\documentclass[twocolumn,showpacs,preprintnumbers,amsmath,amssymb,prl,superscriptaddress]{revtex4-1}
\usepackage{amsfonts}
\usepackage[english]{babel}
\usepackage[T1]{fontenc}
\usepackage{times}
\usepackage{mathrsfs}
\usepackage{graphicx}% Include figure files
\usepackage{dcolumn}% Align table columns on decimal point
\usepackage{bm}% bold math
\usepackage[colorlinks,bookmarks=false,citecolor=blue,linkcolor=red,urlcolor=blue]{hyperref}

\begin{document}
\author{Zhao Liu}
\affiliation{Dahlem Center for Complex Quantum Systems and Institut f\"ur Theoretische Physik, Freie Universit\"at Berlin, Arnimallee 14, 14195 Berlin, Germany}
\affiliation{Department of Electrical Engineering, Princeton University, Princeton, New Jersey 08544, USA}
\author{R. N. Bhatt}
\affiliation{Department of Electrical Engineering, Princeton University, Princeton, New Jersey 08544, USA}

\title{Quantum Entanglement as a Diagnostic of Phase Transitions in Disordered\\ Fractional Quantum Hall Liquids}
\date{\today}

\begin{abstract}
We investigate the disorder-driven phase transition from  a fractional quantum Hall state to an Anderson insulator using quantum entanglement methods. We find that the transition is signaled by a sharp increase in the sensitivity of a suitably averaged entanglement entropy with respect to disorder -- the magnitude of its disorder derivative appears to diverge in the thermodynamic limit. We also study the level statistics of the entanglement spectrum as a function of disorder. However, unlike the dramatic phase-transition signal in the entanglement entropy derivative, we find a gradual reduction of level repulsion only deep in the Anderson insulating phase.
\end{abstract}

\maketitle

{\em Introduction.---}
Following the advances in understanding the fascinatingly complex phase diagram of two-dimensional electron systems in a strong perpendicular magnetic field in the fractional quantum Hall (FQH) regime \cite{haldane95,eddy99,panw,LL3,negativeflux,papic,cfpg,bilayerScott,bilayerLiu,bilayerPeterson,MRpg,nicolaswigner}, there has been intense interest in phase transitions in topological systems \cite{read,fqhedge,wenbook,wangzh,fradkin}. Disorder is a ubiquitous ingredient that may affect, even drive such phase transitions. In fact, it has been understood since the original explanation of the integer QH effect \cite{laughlinIQH} that disorder plays an essential role in providing a finite domain for various QH states. Its presence allows a detailed study of QH plateau transitions in experiments \cite{wei88,song97,li05,li09}, and in numerical studies of noninteracting models \cite{Huckestein90,huo,Lee93,dnsheng98,Slevin09}.

By contrast, there have been relatively few numerical studies of transitions from FQH states driven by disorder. The breaking of spatial symmetries by random disorder makes numerics even more challenging for this many-body problem with an exponentially large Hilbert space. One study \cite{dnsheng,xin} over a decade ago examined the disorder-driven transition from the $f=1/3$ filling FQH state to the insulator using the ground-state Chern number as a diagnostic. It was shown that with increased disorder, the gap characterizing the FQH state collapsed, leading to an insulating phase at large disorder. Further, the calculated disorder dependence of the gap agreed with experiments.

In recent years, topological phases have been characterized by the underlying patterns of quantum entanglement \cite{kitaevgamma,wengamma,eisert2010}. Besides the vast theoretical literature, entanglement has been the subject of recent experiments \cite{greiner} and related proposals \cite{hafezi}. Though many studies have been done for clean FQH states using the concepts of entanglement entropy and the entanglement spectrum (see Refs.~\cite{fqhgammasphere,fqhgammatorus,hli,nicolas09,andreas,ronny10,papic11,PES,zhao12,dubail2012,parsa2012,vidal2013,jain2013,zaletel2013,wei2015,peterson2015,estienne2015} and references therein), there is relatively little corresponding effort \cite{friedman1, friedman2} in the presence of disorder. In this work, we fill this void by studying the ground-state entanglement properties in a disordered $f=1/3$ FQH system, as the disorder strength $W$ is increased. We find that the magnitude of the derivative of a suitably defined entanglement entropy $S$ with respect to $W$ exhibits a sharp peak at a characteristic value of $W = W_c$. The peak increases with system size, and is consistent with a divergence in the thermodynamic limit. We identify this behavior as signaling the FQH-insulator phase transition. Besides being a completely different diagnostic of the phase transition, our method (which only uses periodic boundary conditions) is significantly faster than calculating Chern numbers \cite{dnsheng,xin}, which requires integrating over boundary conditions. We also analyze the entanglement spectrum (ES), and find that its level statistics undergoes an evolution which differs from that of highly excited states studied very recently in the context of many-body localization \cite{yangzc,nicolas,powerlawES}. In our case, the low-energy part of the ES obeys Gaussian Unitary Ensemble (GUE) statistics at low $W$, and changes to Poisson only at very large $W$.

When the entanglement entropy changes its scaling behavior at a phase transition,  e.g., from a volume law to an area law, one can define an ``order parameter'' by dividing by the volume. One thus obtains a zero order parameter on one side of the transition, and finite on the other, just like symmetry-breaking transitions. In our case, however, the entanglement entropy follows an area law on both sides \cite{fqhgammasphere,fqhgammatorus,kun}, so this method does not work. Instead, we analyze our data using finite-size scaling (FSS) ideas (see, e.g., Ref.~\cite{Chandran}), and do obtain a good FSS collapse of the data \cite{newref}. However, the exponent $\nu$ that we obtain for the diverging length scale violates the expected inequality $\nu\geq2/d$ for $d-$dimensional disordered systems \cite{Chayes86,harris,Mott}, a violation also seen in recent studies of many-body localization \cite{pollman, alet}. We provide possible explanations of this result in our concluding remarks.

{\em Model.---}
We consider $N$ interacting electrons in a two-dimensional random potential on a square with periodic boundary conditions (torus geometry). Landau levels are formed in the presence of a perpendicular external magnetic field. We assume that the energy scales of both interaction and disorder are small compared with the Landau level spacing, so we can focus on the lowest Landau level (LLL). The Hamiltonian of the system is $H=\mathcal{P}_{{\rm LLL}}\Big[\sum_{i<j}^N V({\bf r}_i-{\bf r}_j)+\sum_{i=1}^N U({\bf r}_i)\Big]$, where $\mathcal{P}_{{\rm LLL}}$ is the projector to the LLL, and $V({\bf r})$ and $U({\bf r})$ are the interaction and the random potentials, respectively. We consider Gaussian white noise disorder with strength $W$, satisfying $\langle U({\bf r})\rangle=0$ and $\langle U({\bf r})U({\bf r}')\rangle=W^2 \delta({\bf r}-{\bf r}')$. We further suppose that electrons interact via Haldane's pseudopotential \cite{haldaneVm} $V({\bf r})\propto\nabla^2\delta({\bf r})$ \cite{V1_note}.

We consider partially filled LLL at filling $f=N/N_\phi=1/3$, where $N_\phi=3N$ is the number of orbitals in the LLL. In clean samples with $W=0$, Haldane's pseudopotential guarantees that the ground states are exact Laughlin states \cite{laughlin} protected by energy and mobility gaps to excited levels. Increasing disorder gradually closes both gaps \cite{dnsheng,xin}, leading to a phase transition from the topological phase. For $W=\infty$, we recover the noninteracting limit, where extended single-particle states only exist at the center of the LLL band \cite{huo}. Since all single-particle states below the Fermi level at $f=1/3$ are localized, the ground state is an Anderson insulator. In the following, we monitor the ground-state entanglement properties as a function of $W$ \cite{systemsize}.

\begin{figure}
\centerline{\includegraphics[width=0.8\linewidth]{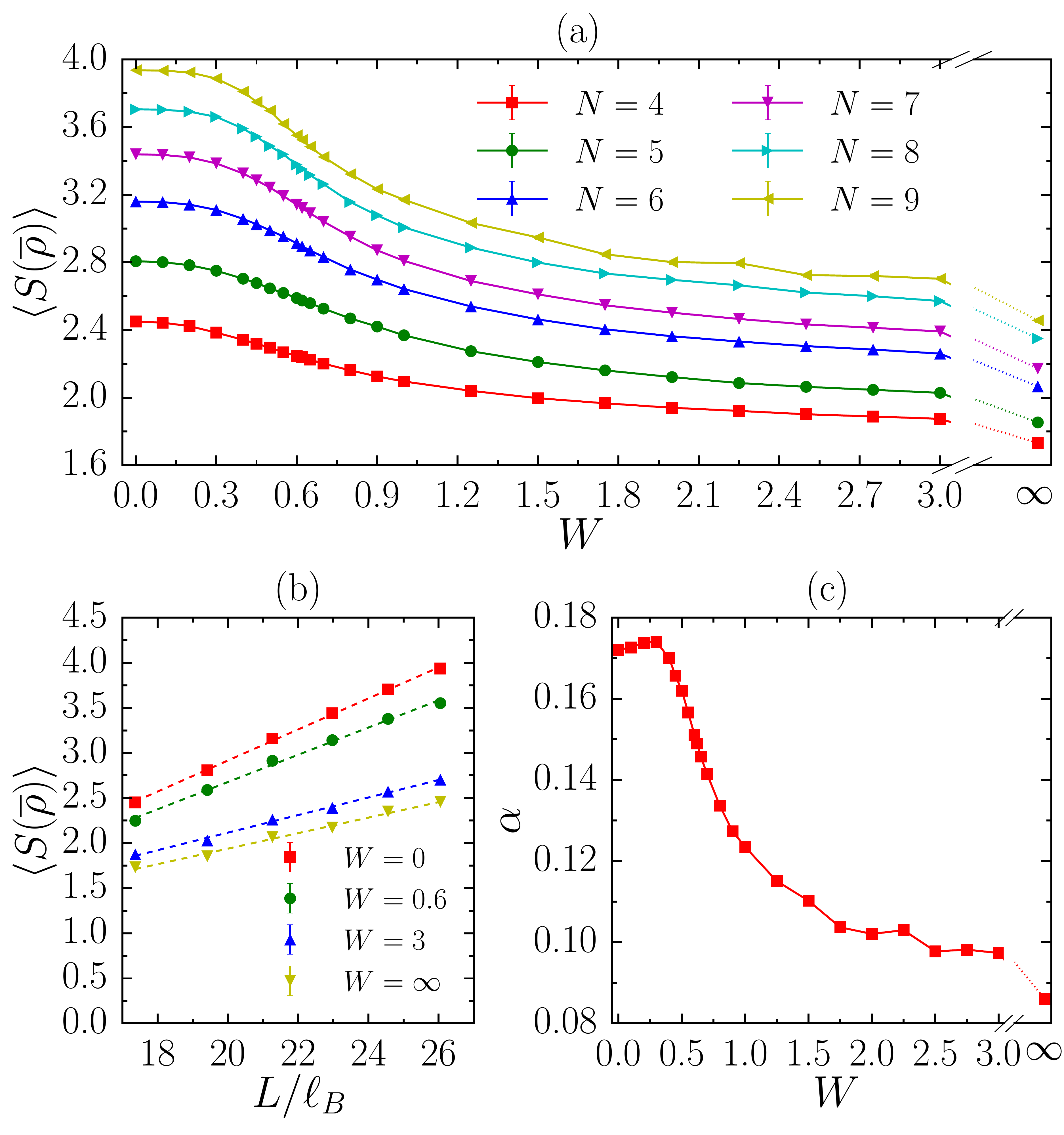}}
\caption{(a) 
$\langle S(\overline{\rho}) \rangle$ versus 
$W$ for $N=4-9$ electrons. (b) $\langle S(\overline{\rho}) \rangle$ versus the cut length $L$ for $N=4-9$ electrons at $W=0,0.6,3$ and $\infty$. The dotted line is obtained by the linear fitting $\langle S(\overline{\rho}) \rangle=\alpha L-\beta$. (c) The entanglement density $\alpha$ versus $W$. We averaged $20000$ samples for $N=4-7$, $5000$ samples for $N=8$, and $800$ samples for $N=9$ electrons. In (a) and (c), we also give the data at $W=\infty$.}
\label{SA}
\end{figure}

{\em Ground-state entanglement entropy.---}
The entanglement in a bipartite system can be measured by the von Neumann entropy of one subsystem. We make two cuts at orbital $0$ and $\lceil N_\phi/2\rfloor-1$, respectively, where $\lceil x\rfloor$ is the integer part of $x$. This procedure divides $N_\phi$ LLL orbitals into subsystems $A$ and $B$ with boundary length $L=2\times\sqrt{2\pi N_\phi}$ (in units of the magnetic length $\ell_B$), consisting of orbitals $0,...,\lceil N_\phi/2\rfloor-1$ and $\lceil N_\phi/2\rfloor,...,N_\phi-1$, respectively. In clean samples, there are three exactly degenerate Laughlin states. Disorder splits them. However, in the topological phase, the degeneracy is recovered in the thermodynamic limit \cite{wen}. Such a topological degeneracy motivates us to consider a ground-state manifold containing the lowest three eigenstates $|\Psi_{i=1,2,3}\rangle$ of the Hamiltonian at any $W$ for consistency, rather than a single eigenstate. Similar to the situation of a single ground state, we define the ground-state entanglement entropy in our case as $S(\rho)=-{\rm Tr}\rho_A\ln\rho_A$, where $\rho_A={\rm Tr}_B\rho$ is the reduced density matrix of part $A$, and $\rho$ is the density matrix describing the ground-state manifold. There are several options to choose $\rho$, for example, by using either $S(\overline{\rho})$ with $\overline{\rho}=\frac{1}{3}\sum_{i=1}^3|\Psi_i\rangle\langle\Psi_i|$ or $\overline{S}=\frac{1}{3}\sum_{i=1}^3 S(|\Psi_i\rangle)$. Both of them include the contributions of all $|\Psi_i\rangle$'s, thus minimizing the finite-size effect. Moreover, they give similar results \cite{supple}, so we adopt the first one in what follows.

At each $W$, we implement different disorder configurations and compute the sample-averaged entanglement entropy $\langle S(\overline{\rho})\rangle$.
The evolution of $\langle S(\overline{\rho})\rangle$ as a function of $W$ is shown in Fig.~\ref{SA}(a) for various system sizes. For a fixed system size, $\langle S(\overline{\rho}) \rangle$ decreases with $W$. However, it increases with the system size at a fixed $W$. $\langle S(\overline{\rho})\rangle$, although likely suffering from finite-size effects for our small sizes, agrees with an area law variation $\langle S(\overline{\rho}) \rangle=\alpha L-\beta$ \cite{kitaevgamma,wengamma,eisert2010} at all $W$'s [Fig.~\ref{SA}(b)]. The extracted entanglement density $\alpha$ starts to drop at $W\approx0.4$, providing a rough estimate for the collapse of the Laughlin phase.

\begin{figure}[b]
\centerline{\includegraphics[width=\linewidth]{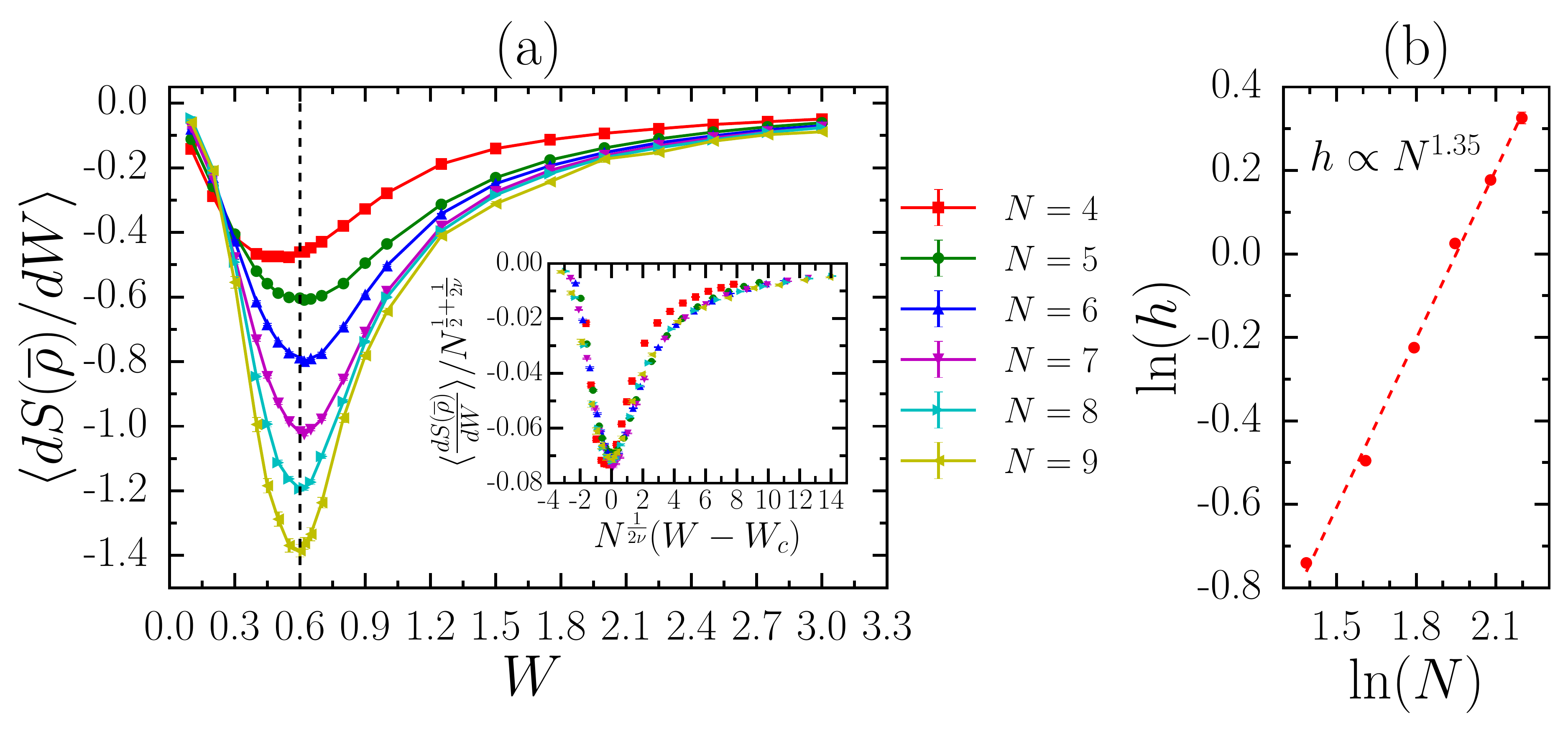}}
\caption{(a) 
$\langle dS(\overline{\rho})/dW\rangle$ versus 
$W$ for $N=4-9$ electrons, replotted in terms of scaled variables $\langle \frac{dS(\overline{\rho})}{dW}\rangle/N^{\frac{1}{2}+\frac{1}{2\nu}}$ and $N^{\frac{1}{2\nu}}(W-W_c)$ in the inset with $W_c=0.6$ and $1/\nu=1.7$. (b) The depth of the minimum $h$ versus $N$ on a double logarithmic scale. The dashed line corresponds to $h\propto N^{1.35}$. We averaged $20000$ samples for $N=4-7$, $5000$ samples for $N=8$, and $800$ samples for $N=9$ electrons.}
\label{dsdw}
\end{figure}

A more precise location of the phase transition can be obtained from the {\it derivative} of the entanglement entropy with respect to the disorder strength, i.e., $dS(\overline{\rho})/dW$. To compute the sample-averaged derivative $\langle dS(\overline{\rho})/dW \rangle$, we average the derivative for each disorder configuration, approximated by $dS(\overline{\rho})/dW=[S(\overline{\rho})|_{W+\Delta W}-S(\overline{\rho})|_W]/\Delta W$ \cite{dsdw_note}. Figure \ref{dsdw}(a) shows $\langle dS(\overline{\rho})/dW \rangle$ as a function of $W$ for various system sizes. All curves exhibit a pronounced minimum at $W_c\approx0.6$, which gets deeper and sharper with increasing system size. A double logarithmic plot of its magnitude $h=|\min\langle dS(\overline{\rho})/dW\rangle|$ versus $N$ [Fig.~\ref{dsdw}(b)] shows that $h\propto N^{1.35}$, consistent with a divergence in the thermodynamic limit. We consider this divergence to be a convincing signature of a phase transition between two phases with different entanglement properties.

Since the entanglement entropy obeys an area law in both phases, for a continuous phase transition, we may expect a scaling behavior for large size $N$ of the form $ S(\overline{\rho}) \propto N^{\frac{1}{2}}f[N^{\frac{1}{2\nu}}(W-W_c)]$, implying $dS(\overline{\rho})/dW\propto N^{\frac{1}{2}+\frac{1}{2\nu}}f'[N^{\frac{1}{2\nu}}(W-W_c)]$. Thus Fig.~\ref{dsdw}(b) implies that $\nu\approx0.6$. In fact, we find that besides the smallest size $N=4$, all the curves in Fig.~\ref{dsdw}(a) collapse onto a single scaled plot [Fig.~\ref{dsdw}(a), inset] for $\nu\approx0.6$.

\begin{figure}
\centerline{\includegraphics[width=0.8\linewidth]{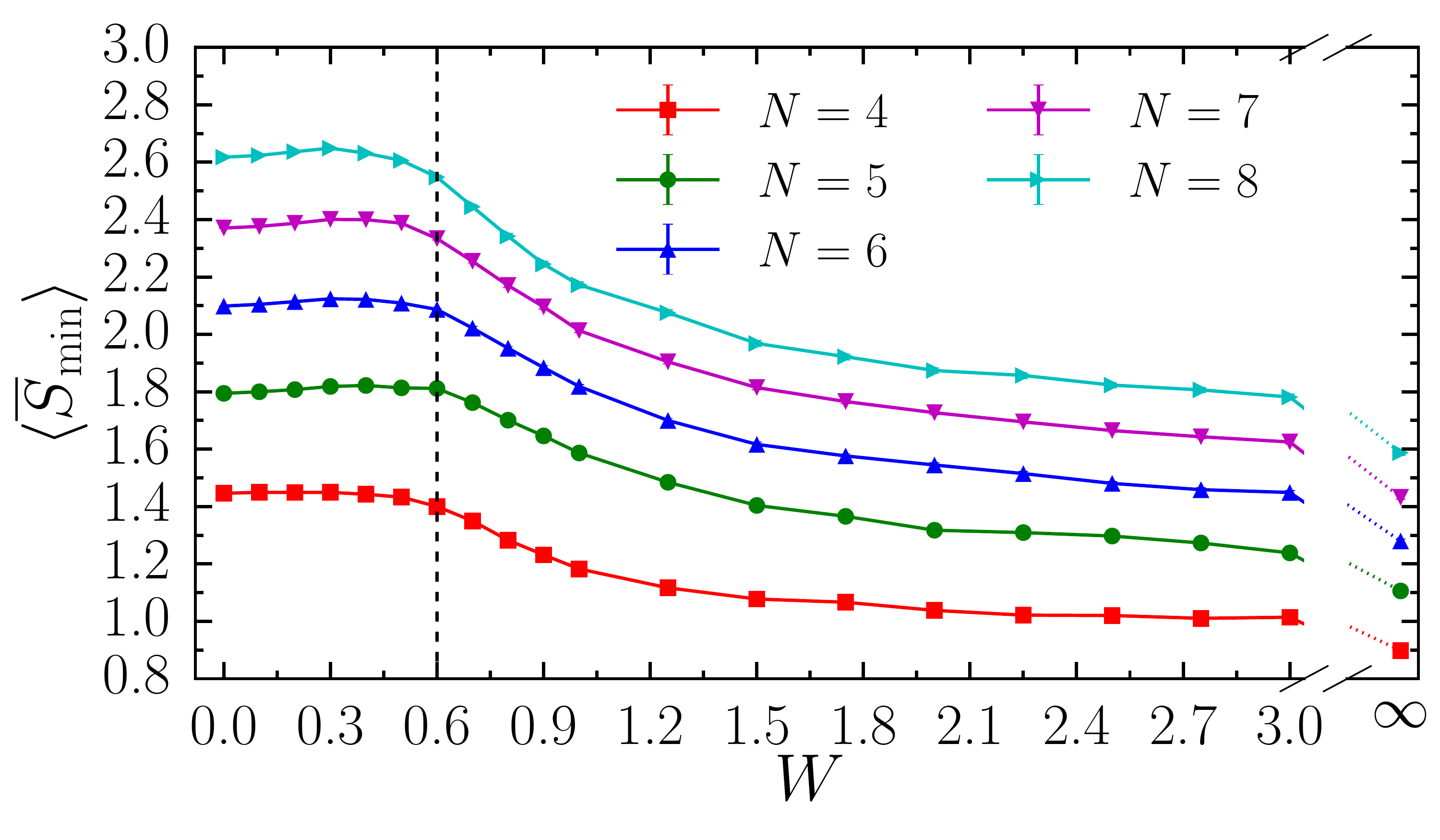}}
\caption{
$\langle \overline{S}_{\min}\rangle$ versus 
$W$ for $N=4-8$ electrons. We averaged $2000$ samples for $N=4-7$ and $1000$ samples for $N=8$ electrons. The data at $W=\infty$ are also given.}
\label{mes}
\end{figure}

The information of ground-state entanglement can also be extracted from minimally entangled states (MES) in the ground-state manifold.
For all superpositions $|\Psi\rangle=\sin\theta_1\sin\theta_2|\Psi_1\rangle+\sin\theta_1\cos\theta_2 e^{i\phi_1}|\Psi_2\rangle+\cos\theta_1 e^{i\phi_2}|\Psi_3\rangle$ with $\theta_1,\theta_2\in[0,\pi/2]$ and $\phi_1,\phi_2\in[0,2\pi)$, the local minima of $S(|\Psi\rangle)$ in the parameter space spanned by $(\theta_1,\theta_2,\phi_1,\phi_2)$ correspond to the MES. In the presence of topological degeneracy between $|\Psi_i\rangle$'s, MES are essential for the extraction of the topological entropy \cite{dong2008,zhangyi} and modular matrices \cite{zhangyi,wei1,wei2}. We numerically search all minimally entangled states $|\Psi_{\min}^i\rangle,i=1,...,M$ in each sample \cite{cleanmes}, and compute their average entanglement entropy $\overline{S}_{\min}=\frac{1}{M}\sum_{i=1}^MS(|\Psi_{\min}^i\rangle)$. In Fig.~\ref{mes}, we show the sample-averaged entropy $\langle \overline{S}_{\min}\rangle$ as a function of $W$. At small disorder, $\langle \overline{S}_{\min}\rangle$ is almost a constant, indicating that the ground-state topological properties are the same as those of Laughlin states in clean samples. $\langle \overline{S}_{\min}\rangle$ starts to drop at $W\approx0.6$, signifying a phase transition. Strikingly, the critical $W$ obtained from the minimally entangled states is the same as that suggested by the singular behavior of $\langle dS(\overline{\rho})/dW \rangle$.

{\em Ground-state entanglement spectrum (ES).---} The spectrum of the reduced density matrix $\rho_A$ usually contains more information than the entanglement entropy, which is a single number obtained from the whole spectrum. Therefore, we now consider the ground-state ES $\{\xi_i\}$ -- the spectrum of $-\ln\rho_A$. Since the number of electrons $N_A$ in part $A$ is still conserved in the presence of disorder, we focus on the ES with $N_A=\lceil N/2\rfloor$, which has the largest rank.

In clean torus samples, the ES levels, when plotted versus their momenta, match the combination of two edge modes of pertinent FQH states \cite{andreas,zliu}. Disorder breaks the translational symmetry; therefore, we expect that ground-state properties will be revealed by the {\it statistics} of the ES \cite{Prodan}. We compute the distribution $P(s)$ of the normalized level spacing $s_n/\langle s_n\rangle$, where $s_n=\xi_n-\xi_{n-1}$ with ${\xi_n}$ sorted in ascending order, as a diagnostic of ES level statistics. To allow for variation between different parts of the ES, we examine level statistics in different windows of $\xi$ in the density of states (DOS) $D(\xi)$. Results from three individual ground states are quite similar \cite{supple}, so we consider $\overline{D}(\xi)$ and $\overline{P}(s)$ averaged over three ground states below. This gives us more statistics in the evaluation. With increasing $W$, ES levels with large $\xi$ suffer from machine precision issues \cite{supple}. Consequently, we limit our discussion to those with $\xi\leq 40$, which are reliable.

\begin{figure}[b]
\centerline{\includegraphics[width=\linewidth]{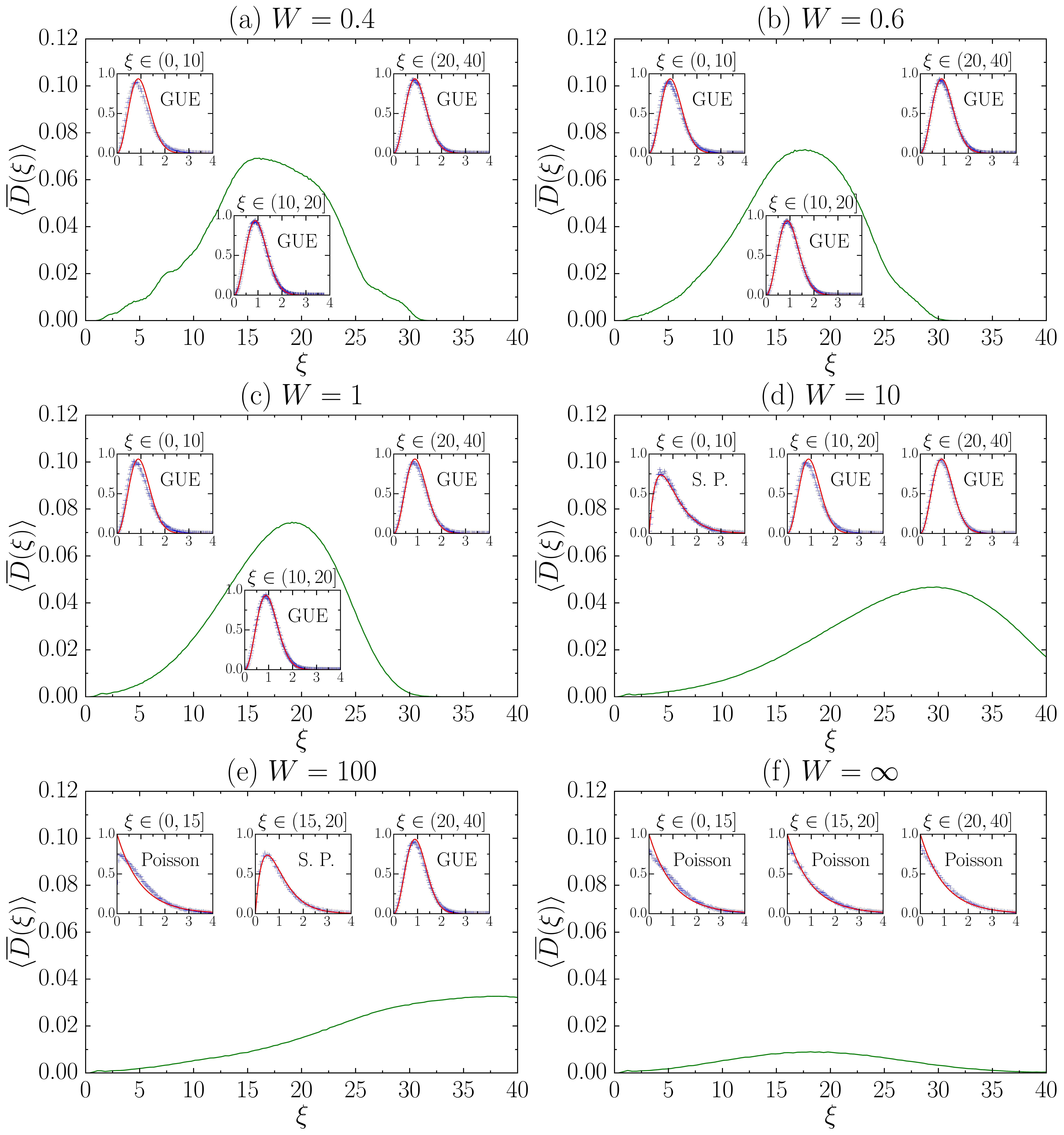}}
\caption{$\overline{P}(s)$ and the sample-averaged DOS $\langle\overline{D}(\xi)\rangle$ 
of the ground-state ES below $\xi=40$ for $N=9$ electrons at (a) $W=0.4$, (b) $W=0.6$, (c) $W=1$, (d) $W=10$, (e) $W=100$, and (f) $W=\infty$. At each $W$, we choose three windows to compute $\overline{P}(s)$, plotted versus $s$ in the insets. The blue crosses correspond to numerical data, while the red lines give the theoretical prediction for the GUE, semi-Poisson (S.~P.) and the Poisson distribution, for which $P(s)=\frac{32}{\pi^2}s^2 e^{-\frac{4}{\pi}s^2}$, $P(s)=4se^{-2s}$, and $P(s)=e^{-s}$, respectively. Data from $800$ realizations of disorder.
}
\label{oes}
\end{figure}

With increasing $W$, we observe the development of localization in the ES. Localization first appears among low ES levels, then expands to higher$-\xi$ levels at larger $W$. In Fig.~\ref{oes}, we show the ES level statistics in three windows \cite{window_note} for $N=9$ electrons at various disorder strengths. At $W=0.4$, we find strong level repulsion governed by GUE in all three windows; thus, the level spacing of the entire ES follows GUE [Fig.~\ref{oes}(a)]. The same ES level statistics is found at $W=0.6$ [Fig.~\ref{oes}(b)] and $W=1$ [Fig.~\ref{oes}(c)]. Only at significantly higher $W=10$ is there a significant change in the level statistics -- while $\overline{P}(s)$ in the two higher$-\xi$ windows still obeys GUE, the level repulsion in $\xi\in(0,10]$ is weaker, where GUE is replaced by a semi-Poisson distribution [Fig.~\ref{oes}(d)]. At $W=100$, the spacing of the low ES levels becomes almost Poissonian without level repulsion, as for a localized system. Simultaneously, the level statistics in the middle window has evolved to semi-Poisson, while GUE is still preserved in the third window [Fig.~\ref{oes}(e)]. Finally, in the noninteracting limit with $W=\infty$, we observe the Poisson distribution for all ES levels below $\xi=40$ [Fig.~\ref{oes}(f)]. It should be emphasized that this evolution of the ES takes place at significantly higher values of $W$ than the ground-state phase transition at $W\approx0.6$ indicated by the entanglement entropy derivative.

We also track the number of reliable ES levels below $\xi=40$ to get more insights about the ES evolution.
We find that more and more ES levels move to the $\xi>40$ region with increasing disorder \cite{n<40}. At $W=\infty$, only a small fraction of ES levels ($18\%$ for $N=9$ and $8\%$ for $N=10$ electrons) forms one peak below $\xi=40$ [Fig.~\ref{oes}(f)], where the Poisson distribution dominates the level statistics. The majority of ES levels have moved to the $\xi>40$ region, which we cannot track due to the machine precision problem. Since ES levels in the rightmost part of the DOS always display GUE [Figs.~\ref{oes}(a)-\ref{oes}(e)], we might expect that the level statistics in the $\xi>40$ region at $W=\infty$ also follows GUE. Since $\langle\overline{D}(\xi)\rangle$ is almost zero at $\xi=40$, this would imply that there is an entanglement gap separating the Poisson part and GUE part for $W=\infty$.

Summarizing the analysis above, we can obtain the following picture of the ES evolution. The entire ES follows GUE at small disorder [Figs.~\ref{oes}(a)-\ref{oes}(c)]. With the increase of $W$, localization is first activated among low ES levels, then propagates towards higher$-\xi$ region, characterized by the evolution of level statistics from GUE to semi-Poisson, then to Poisson distribution [Figs.~\ref{oes}(d)-\ref{oes}(f)]. However, ES levels with GUE still exist, which are located at higher and higher $\xi$ with the expansion of the DOS [Figs.~\ref{oes}(d),\ref{oes}(e)]. 
Finally, in the noninteracting limit, the Poisson part forms a wide and small peak in the DOS [Fig.~\ref{oes}(f)], where only a small fraction of ES levels are located. The majority of ES levels belong to the GUE part with quite high $\xi$ and is separated from the Poisson part by an entanglement gap. We notice that another two-component structure was also observed in the ES of highly excited energy eigenstates, but with different level statistics \cite{yangzc,nicolas}, reflecting the different evolution of the ground state and highly excited states with disorder.

{\em Discussion.---}
In summary, through extensive exact-diagonalization studies of quantum entanglement properties of the $f=1/3$ FQH system with $N=4-9$ electrons on a torus geometry as a function of disorder, we find that the system undergoes a transition from the topological Laughlin state to a (topologically trivial) Anderson insulating state. The phase transition is signaled by a sharp peak (at a characteristic disorder $W_c \approx0.6$) in the magnitude of the disorder derivative of the entanglement entropy, $dS/dW$, which appears to diverge in the thermodynamic limit
(much as thermal transitions are often characterized by a singularity of the specific heat, which is proportional to the temperature derivative of the thermal entropy)
. From the scaling of the magnitude of the peak with size, as well as the entire $dS/dW$ curve as a function of disorder in the vicinity of the transition, we obtain $\nu\approx0.6$, which is very different from the exponent $\nu \approx 2.4 - 2.6$ characterizing plateau transitions \cite{wei88,song97,li05,li09,Huckestein90,huo,Lee93,dnsheng98,Slevin09}. Further, it violates the conventionally accepted bound \cite{Chayes86} for {\em nontopological} transitions of disordered systems in two dimensions, which the result for integer quantum Hall transitions of noninteracting electrons does obey. A possible explanation for this anomalous result could be corrections to finite-size scaling, due to subleading terms in the entanglement entropy. However, in that case, the good finite-size scaling of the data will have to be accepted as being fortuitous. Violation of the bound was also found in recent numerical studies of many-body localization in one dimension \cite{pollman,alet}. Whatever the resolution of this scaling, the location of the transition appears to be quite robust. We find similar results for transitions of Coulomb interacting electrons from other FQH states to the insulator, to be reported in a subsequent publication \cite{zhao-bhatt}. Our results provide motivation for experimental studies of the length exponent characterizing the transition to the insulating state from various FQH states. A study of the universality of the diagonal conductivity at such a transition \cite{shahar} was motivated by theoretical \cite{kivelson} and numerical \cite{huohetzelbhatt} studies. In contrast to the entanglement entropy, the evolution of the level spacing of the entanglement spectrum is found to occur at much larger disorder, over an order of magnitude larger than the critical disorder, showing evidence of level repulsion in the entanglement spectrum until deep in the Anderson insulating phase. %Finally, the very recent experimental study \cite{greiner} and proposal \cite{hafezi} of entanglement properties in cold atom systems forbodes well for the utility of quantum entanglement in studies of phase transitions in condensed matter.%

\acknowledgments
We thank B.~A.~Bernevig, S.~D. Geraedts, F.~D.~M. Haldane, D.~A. Huse, Rahul Nandkishore, Nicolas Regnault, and Xin Wan for helpful discussions. This work was supported by the Department of Energy, Office of Basic Energy Sciences through Grant No.~DE-SC0002140. Z.~L. was also supported by the Dahlem Research School (DRS) Postdoc Fellowship and Alexander von Humboldt Research Fellowship for Postdoctoral Researchers.

\bibliography{disorder}

\begin{thebibliography}{84}%
\makeatletter
\providecommand \@ifxundefined [1]{%
 \@ifx{#1\undefined}
}%
\providecommand \@ifnum [1]{%
 \ifnum #1\expandafter \@firstoftwo
 \else \expandafter \@secondoftwo
 \fi
}%
\providecommand \@ifx [1]{%
 \ifx #1\expandafter \@firstoftwo
 \else \expandafter \@secondoftwo
 \fi
}%
\providecommand \natexlab [1]{#1}%
\providecommand \enquote  [1]{``#1''}%
\providecommand \bibnamefont  [1]{#1}%
\providecommand \bibfnamefont [1]{#1}%
\providecommand \citenamefont [1]{#1}%
\providecommand \href@noop [0]{\@secondoftwo}%
\providecommand \href [0]{\begingroup \@sanitize@url \@href}%
\providecommand \@href[1]{\@@startlink{#1}\@@href}%
\providecommand \@@href[1]{\endgroup#1\@@endlink}%
\providecommand \@sanitize@url [0]{\catcode `\\12\catcode `\$12\catcode
  `\&12\catcode `\#12\catcode `\^12\catcode `\_12\catcode `\%12\relax}%
\providecommand \@@startlink[1]{}%
\providecommand \@@endlink[0]{}%
\providecommand \url  [0]{\begingroup\@sanitize@url \@url }%
\providecommand \@url [1]{\endgroup\@href {#1}{\urlprefix }}%
\providecommand \urlprefix  [0]{URL }%
\providecommand \Eprint [0]{\href }%
\providecommand \doibase [0]{http://dx.doi.org/}%
\providecommand \selectlanguage [0]{\@gobble}%
\providecommand \bibinfo  [0]{\@secondoftwo}%
\providecommand \bibfield  [0]{\@secondoftwo}%
\providecommand \translation [1]{[#1]}%
\providecommand \BibitemOpen [0]{}%
\providecommand \bibitemStop [0]{}%
\providecommand \bibitemNoStop [0]{.\EOS\space}%
\providecommand \EOS [0]{\spacefactor3000\relax}%
\providecommand \BibitemShut  [1]{\csname bibitem#1\endcsname}%
\let\auto@bib@innerbib\@empty
%</preamble>
\bibitem [{\citenamefont {McDonald}\ and\ \citenamefont
  {Haldane}(1996)}]{haldane95}%
  \BibitemOpen
  \bibfield  {author} {\bibinfo {author} {\bibfnamefont {I.~A.}\ \bibnamefont
  {McDonald}}\ and\ \bibinfo {author} {\bibfnamefont {F.~D.~M.}\ \bibnamefont
  {Haldane}},\ }\href {\doibase 10.1103/PhysRevB.53.15845} {\bibfield
  {journal} {\bibinfo  {journal} {Phys. Rev. B}\ }\textbf {\bibinfo {volume}
  {53}},\ \bibinfo {pages} {15845} (\bibinfo {year} {1996})}\BibitemShut
  {NoStop}%
\bibitem [{\citenamefont {Ardonne}\ and\ \citenamefont
  {Schoutens}(1999)}]{eddy99}%
  \BibitemOpen
  \bibfield  {author} {\bibinfo {author} {\bibfnamefont {E.}~\bibnamefont
  {Ardonne}}\ and\ \bibinfo {author} {\bibfnamefont {K.}~\bibnamefont
  {Schoutens}},\ }\href {\doibase 10.1103/PhysRevLett.82.5096} {\bibfield
  {journal} {\bibinfo  {journal} {Phys. Rev. Lett.}\ }\textbf {\bibinfo
  {volume} {82}},\ \bibinfo {pages} {5096} (\bibinfo {year}
  {1999})}\BibitemShut {NoStop}%
\bibitem [{\citenamefont {Pan}\ \emph {et~al.}(2003)\citenamefont {Pan},
  \citenamefont {Stormer}, \citenamefont {Tsui}, \citenamefont {Pfeiffer},
  \citenamefont {Baldwin},\ and\ \citenamefont {West}}]{panw}%
  \BibitemOpen
  \bibfield  {author} {\bibinfo {author} {\bibfnamefont {W.}~\bibnamefont
  {Pan}}, \bibinfo {author} {\bibfnamefont {H.~L.}\ \bibnamefont {Stormer}},
  \bibinfo {author} {\bibfnamefont {D.~C.}\ \bibnamefont {Tsui}}, \bibinfo
  {author} {\bibfnamefont {L.~N.}\ \bibnamefont {Pfeiffer}}, \bibinfo {author}
  {\bibfnamefont {K.~W.}\ \bibnamefont {Baldwin}}, \ and\ \bibinfo {author}
  {\bibfnamefont {K.~W.}\ \bibnamefont {West}},\ }\href {\doibase
  10.1103/PhysRevLett.90.016801} {\bibfield  {journal} {\bibinfo  {journal}
  {Phys. Rev. Lett.}\ }\textbf {\bibinfo {volume} {90}},\ \bibinfo {pages}
  {016801} (\bibinfo {year} {2003})}\BibitemShut {NoStop}%
\bibitem [{\citenamefont {Gervais}\ \emph {et~al.}(2004)\citenamefont
  {Gervais}, \citenamefont {Engel}, \citenamefont {Stormer}, \citenamefont
  {Tsui}, \citenamefont {Baldwin}, \citenamefont {West},\ and\ \citenamefont
  {Pfeiffer}}]{LL3}%
  \BibitemOpen
  \bibfield  {author} {\bibinfo {author} {\bibfnamefont {G.}~\bibnamefont
  {Gervais}}, \bibinfo {author} {\bibfnamefont {L.~W.}\ \bibnamefont {Engel}},
  \bibinfo {author} {\bibfnamefont {H.~L.}\ \bibnamefont {Stormer}}, \bibinfo
  {author} {\bibfnamefont {D.~C.}\ \bibnamefont {Tsui}}, \bibinfo {author}
  {\bibfnamefont {K.~W.}\ \bibnamefont {Baldwin}}, \bibinfo {author}
  {\bibfnamefont {K.~W.}\ \bibnamefont {West}}, \ and\ \bibinfo {author}
  {\bibfnamefont {L.~N.}\ \bibnamefont {Pfeiffer}},\ }\href {\doibase
  10.1103/PhysRevLett.93.266804} {\bibfield  {journal} {\bibinfo  {journal}
  {Phys. Rev. Lett.}\ }\textbf {\bibinfo {volume} {93}},\ \bibinfo {pages}
  {266804} (\bibinfo {year} {2004})}\BibitemShut {NoStop}%
\bibitem [{\citenamefont {Jolicoeur}(2007)}]{negativeflux}%
  \BibitemOpen
  \bibfield  {author} {\bibinfo {author} {\bibfnamefont {T.}~\bibnamefont
  {Jolicoeur}},\ }\href {\doibase 10.1103/PhysRevLett.99.036805} {\bibfield
  {journal} {\bibinfo  {journal} {Phys. Rev. Lett.}\ }\textbf {\bibinfo
  {volume} {99}},\ \bibinfo {pages} {036805} (\bibinfo {year}
  {2007})}\BibitemShut {NoStop}%
\bibitem [{\citenamefont {Papi\ifmmode~\acute{c}\else \'{c}\fi{}}\ \emph
  {et~al.}(2010)\citenamefont {Papi\ifmmode~\acute{c}\else \'{c}\fi{}},
  \citenamefont {Goerbig}, \citenamefont {Regnault},\ and\ \citenamefont
  {Milovanovi\ifmmode~\acute{c}\else \'{c}\fi{}}}]{papic}%
  \BibitemOpen
  \bibfield  {author} {\bibinfo {author} {\bibfnamefont {Z.}~\bibnamefont
  {Papi\ifmmode~\acute{c}\else \'{c}\fi{}}}, \bibinfo {author} {\bibfnamefont
  {M.~O.}\ \bibnamefont {Goerbig}}, \bibinfo {author} {\bibfnamefont
  {N.}~\bibnamefont {Regnault}}, \ and\ \bibinfo {author} {\bibfnamefont
  {M.~V.}\ \bibnamefont {Milovanovi\ifmmode~\acute{c}\else \'{c}\fi{}}},\
  }\href {\doibase 10.1103/PhysRevB.82.075302} {\bibfield  {journal} {\bibinfo
  {journal} {Phys. Rev. B}\ }\textbf {\bibinfo {volume} {82}},\ \bibinfo
  {pages} {075302} (\bibinfo {year} {2010})}\BibitemShut {NoStop}%
\bibitem [{\citenamefont {Archer}\ and\ \citenamefont {Jain}(2013)}]{cfpg}%
  \BibitemOpen
  \bibfield  {author} {\bibinfo {author} {\bibfnamefont {A.~C.}\ \bibnamefont
  {Archer}}\ and\ \bibinfo {author} {\bibfnamefont {J.~K.}\ \bibnamefont
  {Jain}},\ }\href {\doibase 10.1103/PhysRevLett.110.246801} {\bibfield
  {journal} {\bibinfo  {journal} {Phys. Rev. Lett.}\ }\textbf {\bibinfo
  {volume} {110}},\ \bibinfo {pages} {246801} (\bibinfo {year}
  {2013})}\BibitemShut {NoStop}%
\bibitem [{\citenamefont {Geraedts}\ \emph {et~al.}(2015)\citenamefont
  {Geraedts}, \citenamefont {Zaletel}, \citenamefont
  {Papi\ifmmode~\acute{c}\else \'{c}\fi{}},\ and\ \citenamefont
  {Mong}}]{bilayerScott}%
  \BibitemOpen
  \bibfield  {author} {\bibinfo {author} {\bibfnamefont {S.}~\bibnamefont
  {Geraedts}}, \bibinfo {author} {\bibfnamefont {M.~P.}\ \bibnamefont
  {Zaletel}}, \bibinfo {author} {\bibfnamefont {Z.}~\bibnamefont
  {Papi\ifmmode~\acute{c}\else \'{c}\fi{}}}, \ and\ \bibinfo {author}
  {\bibfnamefont {R.~S.~K.}\ \bibnamefont {Mong}},\ }\href {\doibase
  10.1103/PhysRevB.91.205139} {\bibfield  {journal} {\bibinfo  {journal} {Phys.
  Rev. B}\ }\textbf {\bibinfo {volume} {91}},\ \bibinfo {pages} {205139}
  (\bibinfo {year} {2015})}\BibitemShut {NoStop}%
\bibitem [{\citenamefont {Liu}\ \emph {et~al.}(2015)\citenamefont {Liu},
  \citenamefont {Vaezi}, \citenamefont {Lee},\ and\ \citenamefont
  {Kim}}]{bilayerLiu}%
  \BibitemOpen
  \bibfield  {author} {\bibinfo {author} {\bibfnamefont {Z.}~\bibnamefont
  {Liu}}, \bibinfo {author} {\bibfnamefont {A.}~\bibnamefont {Vaezi}}, \bibinfo
  {author} {\bibfnamefont {K.}~\bibnamefont {Lee}}, \ and\ \bibinfo {author}
  {\bibfnamefont {E.-A.}\ \bibnamefont {Kim}},\ }\href {\doibase
  10.1103/PhysRevB.92.081102} {\bibfield  {journal} {\bibinfo  {journal} {Phys.
  Rev. B}\ }\textbf {\bibinfo {volume} {92}},\ \bibinfo {pages} {081102(R)}
  (\bibinfo {year} {2015})}\BibitemShut {NoStop}%
\bibitem [{\citenamefont {Peterson}\ \emph {et~al.}(2015)\citenamefont
  {Peterson}, \citenamefont {Wu}, \citenamefont {Cheng}, \citenamefont
  {Barkeshli}, \citenamefont {Wang},\ and\ \citenamefont
  {Das~Sarma}}]{bilayerPeterson}%
  \BibitemOpen
  \bibfield  {author} {\bibinfo {author} {\bibfnamefont {M.~R.}\ \bibnamefont
  {Peterson}}, \bibinfo {author} {\bibfnamefont {Y.-L.}\ \bibnamefont {Wu}},
  \bibinfo {author} {\bibfnamefont {M.}~\bibnamefont {Cheng}}, \bibinfo
  {author} {\bibfnamefont {M.}~\bibnamefont {Barkeshli}}, \bibinfo {author}
  {\bibfnamefont {Z.}~\bibnamefont {Wang}}, \ and\ \bibinfo {author}
  {\bibfnamefont {S.}~\bibnamefont {Das~Sarma}},\ }\href {\doibase
  10.1103/PhysRevB.92.035103} {\bibfield  {journal} {\bibinfo  {journal} {Phys.
  Rev. B}\ }\textbf {\bibinfo {volume} {92}},\ \bibinfo {pages} {035103}
  (\bibinfo {year} {2015})}\BibitemShut {NoStop}%
\bibitem [{\citenamefont {Pakrouski}\ \emph
  {et~al.}(2015{\natexlab{a}})\citenamefont {Pakrouski}, \citenamefont
  {Peterson}, \citenamefont {Jolicoeur}, \citenamefont {Scarola}, \citenamefont
  {Nayak},\ and\ \citenamefont {Troyer}}]{MRpg}%
  \BibitemOpen
  \bibfield  {author} {\bibinfo {author} {\bibfnamefont {K.}~\bibnamefont
  {Pakrouski}}, \bibinfo {author} {\bibfnamefont {M.~R.}\ \bibnamefont
  {Peterson}}, \bibinfo {author} {\bibfnamefont {T.}~\bibnamefont {Jolicoeur}},
  \bibinfo {author} {\bibfnamefont {V.~W.}\ \bibnamefont {Scarola}}, \bibinfo
  {author} {\bibfnamefont {C.}~\bibnamefont {Nayak}}, \ and\ \bibinfo {author}
  {\bibfnamefont {M.}~\bibnamefont {Troyer}},\ }\href {\doibase
  10.1103/PhysRevX.5.021004} {\bibfield  {journal} {\bibinfo  {journal} {Phys.
  Rev. X}\ }\textbf {\bibinfo {volume} {5}},\ \bibinfo {pages} {021004}
  (\bibinfo {year} {2015}{\natexlab{a}})}\BibitemShut {NoStop}%
\bibitem [{\citenamefont {Thiebaut}\ \emph {et~al.}(2015)\citenamefont
  {Thiebaut}, \citenamefont {Regnault},\ and\ \citenamefont
  {Goerbig}}]{nicolaswigner}%
  \BibitemOpen
  \bibfield  {author} {\bibinfo {author} {\bibfnamefont {N.}~\bibnamefont
  {Thiebaut}}, \bibinfo {author} {\bibfnamefont {N.}~\bibnamefont {Regnault}},
  \ and\ \bibinfo {author} {\bibfnamefont {M.~O.}\ \bibnamefont {Goerbig}},\
  }\href {\doibase 10.1103/PhysRevB.92.245401} {\bibfield  {journal} {\bibinfo
  {journal} {Phys. Rev. B}\ }\textbf {\bibinfo {volume} {92}},\ \bibinfo
  {pages} {245401} (\bibinfo {year} {2015})}\BibitemShut {NoStop}%
\bibitem [{\citenamefont {Read}(1989)}]{read}%
  \BibitemOpen
  \bibfield  {author} {\bibinfo {author} {\bibfnamefont {N.}~\bibnamefont
  {Read}},\ }\href {\doibase 10.1103/PhysRevLett.62.86} {\bibfield  {journal}
  {\bibinfo  {journal} {Phys. Rev. Lett.}\ }\textbf {\bibinfo {volume} {62}},\
  \bibinfo {pages} {86} (\bibinfo {year} {1989})}\BibitemShut {NoStop}%
\bibitem [{\citenamefont {Wen}(1995)}]{fqhedge}%
  \BibitemOpen
  \bibfield  {author} {\bibinfo {author} {\bibfnamefont {X.-G.}\ \bibnamefont
  {Wen}},\ }\href {\doibase 10.1080/00018739500101566} {\bibfield  {journal}
  {\bibinfo  {journal} {Adv. Phys.}\ }\textbf {\bibinfo {volume} {44}},\
  \bibinfo {pages} {405} (\bibinfo {year} {1995})}\BibitemShut {NoStop}%
\bibitem [{\citenamefont {Wen}()}]{wenbook}%
  \BibitemOpen
  \bibfield  {author} {\bibinfo {author} {\bibfnamefont {X.-G.}\ \bibnamefont
  {Wen}},\ }\href@noop {} {\emph {\bibinfo {title} {{Quantum Field Theory of
  Many-Body Systems}}}}\ (\bibinfo  {publisher} {Oxford Univ. Press, Oxford,
  2004})\BibitemShut {NoStop}%
\bibitem [{\citenamefont {Gils}\ \emph {et~al.}(2009)\citenamefont {Gils},
  \citenamefont {Trebst}, \citenamefont {Kitaev}, \citenamefont {Ludwig},
  \citenamefont {Troyer},\ and\ \citenamefont {Wang}}]{wangzh}%
  \BibitemOpen
  \bibfield  {author} {\bibinfo {author} {\bibfnamefont {C.}~\bibnamefont
  {Gils}}, \bibinfo {author} {\bibfnamefont {S.}~\bibnamefont {Trebst}},
  \bibinfo {author} {\bibfnamefont {A.}~\bibnamefont {Kitaev}}, \bibinfo
  {author} {\bibfnamefont {A.~W.~W.}\ \bibnamefont {Ludwig}}, \bibinfo {author}
  {\bibfnamefont {M.}~\bibnamefont {Troyer}}, \ and\ \bibinfo {author}
  {\bibfnamefont {Z.}~\bibnamefont {Wang}},\ }\href {\doibase
  10.1038/nphys1396} {\bibfield  {journal} {\bibinfo  {journal} {Nat. Phys}\
  }\textbf {\bibinfo {volume} {5}},\ \bibinfo {pages} {834} (\bibinfo {year}
  {2009})}\BibitemShut {NoStop}%
\bibitem [{\citenamefont {Samkharadze}\ \emph {et~al.}(2016)\citenamefont
  {Samkharadze}, \citenamefont {Schreiber}, \citenamefont {Gardner},
  \citenamefont {Manfra}, \citenamefont {Fradkin},\ and\ \citenamefont
  {Csathy}}]{fradkin}%
  \BibitemOpen
  \bibfield  {author} {\bibinfo {author} {\bibfnamefont {N.}~\bibnamefont
  {Samkharadze}}, \bibinfo {author} {\bibfnamefont {K.~A.}\ \bibnamefont
  {Schreiber}}, \bibinfo {author} {\bibfnamefont {G.~C.}\ \bibnamefont
  {Gardner}}, \bibinfo {author} {\bibfnamefont {M.~J.}\ \bibnamefont {Manfra}},
  \bibinfo {author} {\bibfnamefont {E.}~\bibnamefont {Fradkin}}, \ and\
  \bibinfo {author} {\bibfnamefont {G.~A.}\ \bibnamefont {Csathy}},\ }\href
  {\doibase 10.1038/nphys3523} {\bibfield  {journal} {\bibinfo  {journal} {Nat.
  Phys}\ }\textbf {\bibinfo {volume} {12}},\ \bibinfo {pages} {191} (\bibinfo
  {year} {2016})}\BibitemShut {NoStop}%
\bibitem [{\citenamefont {Laughlin}(1981)}]{laughlinIQH}%
  \BibitemOpen
  \bibfield  {author} {\bibinfo {author} {\bibfnamefont {R.~B.}\ \bibnamefont
  {Laughlin}},\ }\href {\doibase 10.1103/PhysRevB.23.5632} {\bibfield
  {journal} {\bibinfo  {journal} {Phys. Rev. B}\ }\textbf {\bibinfo {volume}
  {23}},\ \bibinfo {pages} {5632} (\bibinfo {year} {1981})}\BibitemShut
  {NoStop}%
\bibitem [{\citenamefont {Wei}\ \emph {et~al.}(1988)\citenamefont {Wei},
  \citenamefont {Tsui}, \citenamefont {Paalanen},\ and\ \citenamefont
  {Pruisken}}]{wei88}%
  \BibitemOpen
  \bibfield  {author} {\bibinfo {author} {\bibfnamefont {H.~P.}\ \bibnamefont
  {Wei}}, \bibinfo {author} {\bibfnamefont {D.~C.}\ \bibnamefont {Tsui}},
  \bibinfo {author} {\bibfnamefont {M.~A.}\ \bibnamefont {Paalanen}}, \ and\
  \bibinfo {author} {\bibfnamefont {A.~M.~M.}\ \bibnamefont {Pruisken}},\
  }\href {\doibase 10.1103/PhysRevLett.61.1294} {\bibfield  {journal} {\bibinfo
   {journal} {Phys. Rev. Lett.}\ }\textbf {\bibinfo {volume} {61}},\ \bibinfo
  {pages} {1294} (\bibinfo {year} {1988})}\BibitemShut {NoStop}%
\bibitem [{\citenamefont {Song}\ \emph {et~al.}(1997)\citenamefont {Song},
  \citenamefont {Shahar}, \citenamefont {Tsui}, \citenamefont {Xie},\ and\
  \citenamefont {Monroe}}]{song97}%
  \BibitemOpen
  \bibfield  {author} {\bibinfo {author} {\bibfnamefont {S.-H.}\ \bibnamefont
  {Song}}, \bibinfo {author} {\bibfnamefont {D.}~\bibnamefont {Shahar}},
  \bibinfo {author} {\bibfnamefont {D.~C.}\ \bibnamefont {Tsui}}, \bibinfo
  {author} {\bibfnamefont {Y.~H.}\ \bibnamefont {Xie}}, \ and\ \bibinfo
  {author} {\bibfnamefont {D.}~\bibnamefont {Monroe}},\ }\href {\doibase
  10.1103/PhysRevLett.78.2200} {\bibfield  {journal} {\bibinfo  {journal}
  {Phys. Rev. Lett.}\ }\textbf {\bibinfo {volume} {78}},\ \bibinfo {pages}
  {2200} (\bibinfo {year} {1997})}\BibitemShut {NoStop}%
\bibitem [{\citenamefont {Li}\ \emph {et~al.}(2005)\citenamefont {Li},
  \citenamefont {Cs\'athy}, \citenamefont {Tsui}, \citenamefont {Pfeiffer},\
  and\ \citenamefont {West}}]{li05}%
  \BibitemOpen
  \bibfield  {author} {\bibinfo {author} {\bibfnamefont {W.}~\bibnamefont
  {Li}}, \bibinfo {author} {\bibfnamefont {G.~A.}\ \bibnamefont {Cs\'athy}},
  \bibinfo {author} {\bibfnamefont {D.~C.}\ \bibnamefont {Tsui}}, \bibinfo
  {author} {\bibfnamefont {L.~N.}\ \bibnamefont {Pfeiffer}}, \ and\ \bibinfo
  {author} {\bibfnamefont {K.~W.}\ \bibnamefont {West}},\ }\href {\doibase
  10.1103/PhysRevLett.94.206807} {\bibfield  {journal} {\bibinfo  {journal}
  {Phys. Rev. Lett.}\ }\textbf {\bibinfo {volume} {94}},\ \bibinfo {pages}
  {206807} (\bibinfo {year} {2005})}\BibitemShut {NoStop}%
\bibitem [{\citenamefont {Li}\ \emph {et~al.}(2009)\citenamefont {Li},
  \citenamefont {Vicente}, \citenamefont {Xia}, \citenamefont {Pan},
  \citenamefont {Tsui}, \citenamefont {Pfeiffer},\ and\ \citenamefont
  {West}}]{li09}%
  \BibitemOpen
  \bibfield  {author} {\bibinfo {author} {\bibfnamefont {W.}~\bibnamefont
  {Li}}, \bibinfo {author} {\bibfnamefont {C.~L.}\ \bibnamefont {Vicente}},
  \bibinfo {author} {\bibfnamefont {J.~S.}\ \bibnamefont {Xia}}, \bibinfo
  {author} {\bibfnamefont {W.}~\bibnamefont {Pan}}, \bibinfo {author}
  {\bibfnamefont {D.~C.}\ \bibnamefont {Tsui}}, \bibinfo {author}
  {\bibfnamefont {L.~N.}\ \bibnamefont {Pfeiffer}}, \ and\ \bibinfo {author}
  {\bibfnamefont {K.~W.}\ \bibnamefont {West}},\ }\href {\doibase
  10.1103/PhysRevLett.102.216801} {\bibfield  {journal} {\bibinfo  {journal}
  {Phys. Rev. Lett.}\ }\textbf {\bibinfo {volume} {102}},\ \bibinfo {pages}
  {216801} (\bibinfo {year} {2009})}\BibitemShut {NoStop}%
\bibitem [{\citenamefont {Huckestein}\ and\ \citenamefont
  {Kramer}(1990)}]{Huckestein90}%
  \BibitemOpen
  \bibfield  {author} {\bibinfo {author} {\bibfnamefont {B.}~\bibnamefont
  {Huckestein}}\ and\ \bibinfo {author} {\bibfnamefont {B.}~\bibnamefont
  {Kramer}},\ }\href {\doibase 10.1103/PhysRevLett.64.1437} {\bibfield
  {journal} {\bibinfo  {journal} {Phys. Rev. Lett.}\ }\textbf {\bibinfo
  {volume} {64}},\ \bibinfo {pages} {1437} (\bibinfo {year}
  {1990})}\BibitemShut {NoStop}%
\bibitem [{\citenamefont {Huo}\ and\ \citenamefont {Bhatt}(1992)}]{huo}%
  \BibitemOpen
  \bibfield  {author} {\bibinfo {author} {\bibfnamefont {Y.}~\bibnamefont
  {Huo}}\ and\ \bibinfo {author} {\bibfnamefont {R.~N.}\ \bibnamefont
  {Bhatt}},\ }\href {\doibase 10.1103/PhysRevLett.68.1375} {\bibfield
  {journal} {\bibinfo  {journal} {Phys. Rev. Lett.}\ }\textbf {\bibinfo
  {volume} {68}},\ \bibinfo {pages} {1375} (\bibinfo {year}
  {1992})}\BibitemShut {NoStop}%
\bibitem [{\citenamefont {Lee}\ \emph {et~al.}(1993)\citenamefont {Lee},
  \citenamefont {Wang},\ and\ \citenamefont {Kivelson}}]{Lee93}%
  \BibitemOpen
  \bibfield  {author} {\bibinfo {author} {\bibfnamefont {D.-H.}\ \bibnamefont
  {Lee}}, \bibinfo {author} {\bibfnamefont {Z.}~\bibnamefont {Wang}}, \ and\
  \bibinfo {author} {\bibfnamefont {S.}~\bibnamefont {Kivelson}},\ }\href
  {\doibase 10.1103/PhysRevLett.70.4130} {\bibfield  {journal} {\bibinfo
  {journal} {Phys. Rev. Lett.}\ }\textbf {\bibinfo {volume} {70}},\ \bibinfo
  {pages} {4130} (\bibinfo {year} {1993})}\BibitemShut {NoStop}%
\bibitem [{\citenamefont {Sheng}\ and\ \citenamefont {Weng}(1998)}]{dnsheng98}%
  \BibitemOpen
  \bibfield  {author} {\bibinfo {author} {\bibfnamefont {D.~N.}\ \bibnamefont
  {Sheng}}\ and\ \bibinfo {author} {\bibfnamefont {Z.~Y.}\ \bibnamefont
  {Weng}},\ }\href {\doibase 10.1103/PhysRevLett.80.580} {\bibfield  {journal}
  {\bibinfo  {journal} {Phys. Rev. Lett.}\ }\textbf {\bibinfo {volume} {80}},\
  \bibinfo {pages} {580} (\bibinfo {year} {1998})}\BibitemShut {NoStop}%
\bibitem [{\citenamefont {Slevin}\ and\ \citenamefont
  {Ohtsuki}(2009)}]{Slevin09}%
  \BibitemOpen
  \bibfield  {author} {\bibinfo {author} {\bibfnamefont {K.}~\bibnamefont
  {Slevin}}\ and\ \bibinfo {author} {\bibfnamefont {T.}~\bibnamefont
  {Ohtsuki}},\ }\href {\doibase 10.1103/PhysRevB.80.041304} {\bibfield
  {journal} {\bibinfo  {journal} {Phys. Rev. B}\ }\textbf {\bibinfo {volume}
  {80}},\ \bibinfo {pages} {041304(R)} (\bibinfo {year} {2009})}\BibitemShut
  {NoStop}%
\bibitem [{\citenamefont {Sheng}\ \emph {et~al.}(2003)\citenamefont {Sheng},
  \citenamefont {Wan}, \citenamefont {Rezayi}, \citenamefont {Yang},
  \citenamefont {Bhatt},\ and\ \citenamefont {Haldane}}]{dnsheng}%
  \BibitemOpen
  \bibfield  {author} {\bibinfo {author} {\bibfnamefont {D.~N.}\ \bibnamefont
  {Sheng}}, \bibinfo {author} {\bibfnamefont {X.}~\bibnamefont {Wan}}, \bibinfo
  {author} {\bibfnamefont {E.~H.}\ \bibnamefont {Rezayi}}, \bibinfo {author}
  {\bibfnamefont {K.}~\bibnamefont {Yang}}, \bibinfo {author} {\bibfnamefont
  {R.~N.}\ \bibnamefont {Bhatt}}, \ and\ \bibinfo {author} {\bibfnamefont
  {F.~D.~M.}\ \bibnamefont {Haldane}},\ }\href {\doibase
  10.1103/PhysRevLett.90.256802} {\bibfield  {journal} {\bibinfo  {journal}
  {Phys. Rev. Lett.}\ }\textbf {\bibinfo {volume} {90}},\ \bibinfo {pages}
  {256802} (\bibinfo {year} {2003})}\BibitemShut {NoStop}%
\bibitem [{\citenamefont {Wan}\ \emph {et~al.}(2005)\citenamefont {Wan},
  \citenamefont {Sheng}, \citenamefont {Rezayi}, \citenamefont {Yang},
  \citenamefont {Bhatt},\ and\ \citenamefont {Haldane}}]{xin}%
  \BibitemOpen
  \bibfield  {author} {\bibinfo {author} {\bibfnamefont {X.}~\bibnamefont
  {Wan}}, \bibinfo {author} {\bibfnamefont {D.~N.}\ \bibnamefont {Sheng}},
  \bibinfo {author} {\bibfnamefont {E.~H.}\ \bibnamefont {Rezayi}}, \bibinfo
  {author} {\bibfnamefont {K.}~\bibnamefont {Yang}}, \bibinfo {author}
  {\bibfnamefont {R.~N.}\ \bibnamefont {Bhatt}}, \ and\ \bibinfo {author}
  {\bibfnamefont {F.~D.~M.}\ \bibnamefont {Haldane}},\ }\href {\doibase
  10.1103/PhysRevB.72.075325} {\bibfield  {journal} {\bibinfo  {journal} {Phys.
  Rev. B}\ }\textbf {\bibinfo {volume} {72}},\ \bibinfo {pages} {075325}
  (\bibinfo {year} {2005})}\BibitemShut {NoStop}%
\bibitem [{\citenamefont {Kitaev}\ and\ \citenamefont
  {Preskill}(2006)}]{kitaevgamma}%
  \BibitemOpen
  \bibfield  {author} {\bibinfo {author} {\bibfnamefont {A.}~\bibnamefont
  {Kitaev}}\ and\ \bibinfo {author} {\bibfnamefont {J.}~\bibnamefont
  {Preskill}},\ }\href {\doibase 10.1103/PhysRevLett.96.110404} {\bibfield
  {journal} {\bibinfo  {journal} {Phys. Rev. Lett.}\ }\textbf {\bibinfo
  {volume} {96}},\ \bibinfo {pages} {110404} (\bibinfo {year}
  {2006})}\BibitemShut {NoStop}%
\bibitem [{\citenamefont {Levin}\ and\ \citenamefont {Wen}(2006)}]{wengamma}%
  \BibitemOpen
  \bibfield  {author} {\bibinfo {author} {\bibfnamefont {M.}~\bibnamefont
  {Levin}}\ and\ \bibinfo {author} {\bibfnamefont {X.-G.}\ \bibnamefont
  {Wen}},\ }\href {\doibase 10.1103/PhysRevLett.96.110405} {\bibfield
  {journal} {\bibinfo  {journal} {Phys. Rev. Lett.}\ }\textbf {\bibinfo
  {volume} {96}},\ \bibinfo {pages} {110405} (\bibinfo {year}
  {2006})}\BibitemShut {NoStop}%
\bibitem [{\citenamefont {Eisert}\ \emph {et~al.}(2010)\citenamefont {Eisert},
  \citenamefont {Cramer},\ and\ \citenamefont {Plenio}}]{eisert2010}%
  \BibitemOpen
  \bibfield  {author} {\bibinfo {author} {\bibfnamefont {J.}~\bibnamefont
  {Eisert}}, \bibinfo {author} {\bibfnamefont {M.}~\bibnamefont {Cramer}}, \
  and\ \bibinfo {author} {\bibfnamefont {M.~B.}\ \bibnamefont {Plenio}},\
  }\href {\doibase 10.1103/RevModPhys.82.277} {\bibfield  {journal} {\bibinfo
  {journal} {Rev. Mod. Phys.}\ }\textbf {\bibinfo {volume} {82}},\ \bibinfo
  {pages} {277} (\bibinfo {year} {2010})}\BibitemShut {NoStop}%
\bibitem [{\citenamefont {Islam}\ \emph {et~al.}(2015)\citenamefont {Islam},
  \citenamefont {Ma}, \citenamefont {Preiss}, \citenamefont {Eric~Tai},
  \citenamefont {Lukin}, \citenamefont {Rispoli},\ and\ \citenamefont
  {Greiner}}]{greiner}%
  \BibitemOpen
  \bibfield  {author} {\bibinfo {author} {\bibfnamefont {R.}~\bibnamefont
  {Islam}}, \bibinfo {author} {\bibfnamefont {R.}~\bibnamefont {Ma}}, \bibinfo
  {author} {\bibfnamefont {P.~M.}\ \bibnamefont {Preiss}}, \bibinfo {author}
  {\bibfnamefont {M.}~\bibnamefont {Eric~Tai}}, \bibinfo {author}
  {\bibfnamefont {A.}~\bibnamefont {Lukin}}, \bibinfo {author} {\bibfnamefont
  {M.}~\bibnamefont {Rispoli}}, \ and\ \bibinfo {author} {\bibfnamefont
  {M.}~\bibnamefont {Greiner}},\ }\href {\doibase 10.1038/nature15750}
  {\bibfield  {journal} {\bibinfo  {journal} {Nature (London)}\ }\textbf
  {\bibinfo {volume} {528}},\ \bibinfo {pages} {77} (\bibinfo {year}
  {2015})}\BibitemShut {NoStop}%
\bibitem [{\citenamefont {{Pichler}}\ \emph {et~al.}()\citenamefont
  {{Pichler}}, \citenamefont {{Zhu}}, \citenamefont {{Seif}}, \citenamefont
  {{Zoller}},\ and\ \citenamefont {{Hafezi}}}]{hafezi}%
  \BibitemOpen
  \bibfield  {author} {\bibinfo {author} {\bibfnamefont {H.}~\bibnamefont
  {{Pichler}}}, \bibinfo {author} {\bibfnamefont {G.}~\bibnamefont {{Zhu}}},
  \bibinfo {author} {\bibfnamefont {A.}~\bibnamefont {{Seif}}}, \bibinfo
  {author} {\bibfnamefont {P.}~\bibnamefont {{Zoller}}}, \ and\ \bibinfo
  {author} {\bibfnamefont {M.}~\bibnamefont {{Hafezi}}},\ }\href@noop {} {\
  }\Eprint {http://arxiv.org/abs/1605.08624} {arXiv:1605.08624} \BibitemShut
  {NoStop}%
\bibitem [{\citenamefont {Haque}\ \emph {et~al.}(2007)\citenamefont {Haque},
  \citenamefont {Zozulya},\ and\ \citenamefont {Schoutens}}]{fqhgammasphere}%
  \BibitemOpen
  \bibfield  {author} {\bibinfo {author} {\bibfnamefont {M.}~\bibnamefont
  {Haque}}, \bibinfo {author} {\bibfnamefont {O.}~\bibnamefont {Zozulya}}, \
  and\ \bibinfo {author} {\bibfnamefont {K.}~\bibnamefont {Schoutens}},\ }\href
  {\doibase 10.1103/PhysRevLett.98.060401} {\bibfield  {journal} {\bibinfo
  {journal} {Phys. Rev. Lett.}\ }\textbf {\bibinfo {volume} {98}},\ \bibinfo
  {pages} {060401} (\bibinfo {year} {2007})}\BibitemShut {NoStop}%
\bibitem [{\citenamefont {L\"auchli}\ \emph
  {et~al.}(2010{\natexlab{a}})\citenamefont {L\"auchli}, \citenamefont
  {Bergholtz},\ and\ \citenamefont {Haque}}]{fqhgammatorus}%
  \BibitemOpen
  \bibfield  {author} {\bibinfo {author} {\bibfnamefont {A.~M.}\ \bibnamefont
  {L\"auchli}}, \bibinfo {author} {\bibfnamefont {E.~J.}\ \bibnamefont
  {Bergholtz}}, \ and\ \bibinfo {author} {\bibfnamefont {M.}~\bibnamefont
  {Haque}},\ }\href {http://stacks.iop.org/1367-2630/12/i=7/a=075004}
  {\bibfield  {journal} {\bibinfo  {journal} {New J. Phys.}\ }\textbf {\bibinfo
  {volume} {12}},\ \bibinfo {pages} {075004} (\bibinfo {year}
  {2010}{\natexlab{a}})}\BibitemShut {NoStop}%
\bibitem [{\citenamefont {Li}\ and\ \citenamefont {Haldane}(2008)}]{hli}%
  \BibitemOpen
  \bibfield  {author} {\bibinfo {author} {\bibfnamefont {H.}~\bibnamefont
  {Li}}\ and\ \bibinfo {author} {\bibfnamefont {F.~D.~M.}\ \bibnamefont
  {Haldane}},\ }\href {\doibase 10.1103/PhysRevLett.101.010504} {\bibfield
  {journal} {\bibinfo  {journal} {Phys. Rev. Lett.}\ }\textbf {\bibinfo
  {volume} {101}},\ \bibinfo {pages} {010504} (\bibinfo {year}
  {2008})}\BibitemShut {NoStop}%
\bibitem [{\citenamefont {Regnault}\ \emph {et~al.}(2009)\citenamefont
  {Regnault}, \citenamefont {Bernevig},\ and\ \citenamefont
  {Haldane}}]{nicolas09}%
  \BibitemOpen
  \bibfield  {author} {\bibinfo {author} {\bibfnamefont {N.}~\bibnamefont
  {Regnault}}, \bibinfo {author} {\bibfnamefont {B.~A.}\ \bibnamefont
  {Bernevig}}, \ and\ \bibinfo {author} {\bibfnamefont {F.~D.~M.}\ \bibnamefont
  {Haldane}},\ }\href {\doibase 10.1103/PhysRevLett.103.016801} {\bibfield
  {journal} {\bibinfo  {journal} {Phys. Rev. Lett.}\ }\textbf {\bibinfo
  {volume} {103}},\ \bibinfo {pages} {016801} (\bibinfo {year}
  {2009})}\BibitemShut {NoStop}%
\bibitem [{\citenamefont {L\"auchli}\ \emph
  {et~al.}(2010{\natexlab{b}})\citenamefont {L\"auchli}, \citenamefont
  {Bergholtz}, \citenamefont {Suorsa},\ and\ \citenamefont {Haque}}]{andreas}%
  \BibitemOpen
  \bibfield  {author} {\bibinfo {author} {\bibfnamefont {A.~M.}\ \bibnamefont
  {L\"auchli}}, \bibinfo {author} {\bibfnamefont {E.~J.}\ \bibnamefont
  {Bergholtz}}, \bibinfo {author} {\bibfnamefont {J.}~\bibnamefont {Suorsa}}, \
  and\ \bibinfo {author} {\bibfnamefont {M.}~\bibnamefont {Haque}},\ }\href
  {\doibase 10.1103/PhysRevLett.104.156404} {\bibfield  {journal} {\bibinfo
  {journal} {Phys. Rev. Lett.}\ }\textbf {\bibinfo {volume} {104}},\ \bibinfo
  {pages} {156404} (\bibinfo {year} {2010}{\natexlab{b}})}\BibitemShut
  {NoStop}%
\bibitem [{\citenamefont {Thomale}\ \emph {et~al.}(2010)\citenamefont
  {Thomale}, \citenamefont {Sterdyniak}, \citenamefont {Regnault},\ and\
  \citenamefont {Bernevig}}]{ronny10}%
  \BibitemOpen
  \bibfield  {author} {\bibinfo {author} {\bibfnamefont {R.}~\bibnamefont
  {Thomale}}, \bibinfo {author} {\bibfnamefont {A.}~\bibnamefont {Sterdyniak}},
  \bibinfo {author} {\bibfnamefont {N.}~\bibnamefont {Regnault}}, \ and\
  \bibinfo {author} {\bibfnamefont {B.~A.}\ \bibnamefont {Bernevig}},\ }\href
  {\doibase 10.1103/PhysRevLett.104.180502} {\bibfield  {journal} {\bibinfo
  {journal} {Phys. Rev. Lett.}\ }\textbf {\bibinfo {volume} {104}},\ \bibinfo
  {pages} {180502} (\bibinfo {year} {2010})}\BibitemShut {NoStop}%
\bibitem [{\citenamefont {Papi\ifmmode~\acute{c}\else \'{c}\fi{}}\ \emph
  {et~al.}(2011)\citenamefont {Papi\ifmmode~\acute{c}\else \'{c}\fi{}},
  \citenamefont {Bernevig},\ and\ \citenamefont {Regnault}}]{papic11}%
  \BibitemOpen
  \bibfield  {author} {\bibinfo {author} {\bibfnamefont {Z.}~\bibnamefont
  {Papi\ifmmode~\acute{c}\else \'{c}\fi{}}}, \bibinfo {author} {\bibfnamefont
  {B.~A.}\ \bibnamefont {Bernevig}}, \ and\ \bibinfo {author} {\bibfnamefont
  {N.}~\bibnamefont {Regnault}},\ }\href {\doibase
  10.1103/PhysRevLett.106.056801} {\bibfield  {journal} {\bibinfo  {journal}
  {Phys. Rev. Lett.}\ }\textbf {\bibinfo {volume} {106}},\ \bibinfo {pages}
  {056801} (\bibinfo {year} {2011})}\BibitemShut {NoStop}%
\bibitem [{\citenamefont {Sterdyniak}\ \emph {et~al.}(2011)\citenamefont
  {Sterdyniak}, \citenamefont {Regnault},\ and\ \citenamefont
  {Bernevig}}]{PES}%
  \BibitemOpen
  \bibfield  {author} {\bibinfo {author} {\bibfnamefont {A.}~\bibnamefont
  {Sterdyniak}}, \bibinfo {author} {\bibfnamefont {N.}~\bibnamefont
  {Regnault}}, \ and\ \bibinfo {author} {\bibfnamefont {B.~A.}\ \bibnamefont
  {Bernevig}},\ }\href {\doibase 10.1103/PhysRevLett.106.100405} {\bibfield
  {journal} {\bibinfo  {journal} {Phys. Rev. Lett.}\ }\textbf {\bibinfo
  {volume} {106}},\ \bibinfo {pages} {100405} (\bibinfo {year}
  {2011})}\BibitemShut {NoStop}%
\bibitem [{\citenamefont {Liu}\ \emph {et~al.}(2012{\natexlab{a}})\citenamefont
  {Liu}, \citenamefont {Bergholtz}, \citenamefont {Fan},\ and\ \citenamefont
  {L\"auchli}}]{zhao12}%
  \BibitemOpen
  \bibfield  {author} {\bibinfo {author} {\bibfnamefont {Z.}~\bibnamefont
  {Liu}}, \bibinfo {author} {\bibfnamefont {E.~J.}\ \bibnamefont {Bergholtz}},
  \bibinfo {author} {\bibfnamefont {H.}~\bibnamefont {Fan}}, \ and\ \bibinfo
  {author} {\bibfnamefont {A.~M.}\ \bibnamefont {L\"auchli}},\ }\href {\doibase
  10.1103/PhysRevB.85.045119} {\bibfield  {journal} {\bibinfo  {journal} {Phys.
  Rev. B}\ }\textbf {\bibinfo {volume} {85}},\ \bibinfo {pages} {045119}
  (\bibinfo {year} {2012}{\natexlab{a}})}\BibitemShut {NoStop}%
\bibitem [{\citenamefont {Dubail}\ \emph {et~al.}(2012)\citenamefont {Dubail},
  \citenamefont {Read},\ and\ \citenamefont {Rezayi}}]{dubail2012}%
  \BibitemOpen
  \bibfield  {author} {\bibinfo {author} {\bibfnamefont {J.}~\bibnamefont
  {Dubail}}, \bibinfo {author} {\bibfnamefont {N.}~\bibnamefont {Read}}, \ and\
  \bibinfo {author} {\bibfnamefont {E.~H.}\ \bibnamefont {Rezayi}},\ }\href
  {\doibase 10.1103/PhysRevB.85.115321} {\bibfield  {journal} {\bibinfo
  {journal} {Phys. Rev. B}\ }\textbf {\bibinfo {volume} {85}},\ \bibinfo
  {pages} {115321} (\bibinfo {year} {2012})}\BibitemShut {NoStop}%
\bibitem [{\citenamefont {Sterdyniak}\ \emph {et~al.}(2012)\citenamefont
  {Sterdyniak}, \citenamefont {Chandran}, \citenamefont {Regnault},
  \citenamefont {Bernevig},\ and\ \citenamefont {Bonderson}}]{parsa2012}%
  \BibitemOpen
  \bibfield  {author} {\bibinfo {author} {\bibfnamefont {A.}~\bibnamefont
  {Sterdyniak}}, \bibinfo {author} {\bibfnamefont {A.}~\bibnamefont
  {Chandran}}, \bibinfo {author} {\bibfnamefont {N.}~\bibnamefont {Regnault}},
  \bibinfo {author} {\bibfnamefont {B.~A.}\ \bibnamefont {Bernevig}}, \ and\
  \bibinfo {author} {\bibfnamefont {P.}~\bibnamefont {Bonderson}},\ }\href
  {\doibase 10.1103/PhysRevB.85.125308} {\bibfield  {journal} {\bibinfo
  {journal} {Phys. Rev. B}\ }\textbf {\bibinfo {volume} {85}},\ \bibinfo
  {pages} {125308} (\bibinfo {year} {2012})}\BibitemShut {NoStop}%
\bibitem [{\citenamefont {Cincio}\ and\ \citenamefont
  {Vidal}(2013)}]{vidal2013}%
  \BibitemOpen
  \bibfield  {author} {\bibinfo {author} {\bibfnamefont {L.}~\bibnamefont
  {Cincio}}\ and\ \bibinfo {author} {\bibfnamefont {G.}~\bibnamefont {Vidal}},\
  }\href {\doibase 10.1103/PhysRevLett.110.067208} {\bibfield  {journal}
  {\bibinfo  {journal} {Phys. Rev. Lett.}\ }\textbf {\bibinfo {volume} {110}},\
  \bibinfo {pages} {067208} (\bibinfo {year} {2013})}\BibitemShut {NoStop}%
\bibitem [{\citenamefont {Balram}\ \emph {et~al.}(2013)\citenamefont {Balram},
  \citenamefont {Wu}, \citenamefont {Sreejith}, \citenamefont {W\'ojs},\ and\
  \citenamefont {Jain}}]{jain2013}%
  \BibitemOpen
  \bibfield  {author} {\bibinfo {author} {\bibfnamefont {A.~C.}\ \bibnamefont
  {Balram}}, \bibinfo {author} {\bibfnamefont {Y.-H.}\ \bibnamefont {Wu}},
  \bibinfo {author} {\bibfnamefont {G.~J.}\ \bibnamefont {Sreejith}}, \bibinfo
  {author} {\bibfnamefont {A.}~\bibnamefont {W\'ojs}}, \ and\ \bibinfo {author}
  {\bibfnamefont {J.~K.}\ \bibnamefont {Jain}},\ }\href {\doibase
  10.1103/PhysRevLett.110.186801} {\bibfield  {journal} {\bibinfo  {journal}
  {Phys. Rev. Lett.}\ }\textbf {\bibinfo {volume} {110}},\ \bibinfo {pages}
  {186801} (\bibinfo {year} {2013})}\BibitemShut {NoStop}%
\bibitem [{\citenamefont {Zaletel}\ \emph {et~al.}(2013)\citenamefont
  {Zaletel}, \citenamefont {Mong},\ and\ \citenamefont
  {Pollmann}}]{zaletel2013}%
  \BibitemOpen
  \bibfield  {author} {\bibinfo {author} {\bibfnamefont {M.~P.}\ \bibnamefont
  {Zaletel}}, \bibinfo {author} {\bibfnamefont {R.~S.~K.}\ \bibnamefont
  {Mong}}, \ and\ \bibinfo {author} {\bibfnamefont {F.}~\bibnamefont
  {Pollmann}},\ }\href {\doibase 10.1103/PhysRevLett.110.236801} {\bibfield
  {journal} {\bibinfo  {journal} {Phys. Rev. Lett.}\ }\textbf {\bibinfo
  {volume} {110}},\ \bibinfo {pages} {236801} (\bibinfo {year}
  {2013})}\BibitemShut {NoStop}%
\bibitem [{\citenamefont {Zhu}\ \emph {et~al.}(2015)\citenamefont {Zhu},
  \citenamefont {Gong}, \citenamefont {Haldane},\ and\ \citenamefont
  {Sheng}}]{wei2015}%
  \BibitemOpen
  \bibfield  {author} {\bibinfo {author} {\bibfnamefont {W.}~\bibnamefont
  {Zhu}}, \bibinfo {author} {\bibfnamefont {S.~S.}\ \bibnamefont {Gong}},
  \bibinfo {author} {\bibfnamefont {F.~D.~M.}\ \bibnamefont {Haldane}}, \ and\
  \bibinfo {author} {\bibfnamefont {D.~N.}\ \bibnamefont {Sheng}},\ }\href
  {\doibase 10.1103/PhysRevLett.115.126805} {\bibfield  {journal} {\bibinfo
  {journal} {Phys. Rev. Lett.}\ }\textbf {\bibinfo {volume} {115}},\ \bibinfo
  {pages} {126805} (\bibinfo {year} {2015})}\BibitemShut {NoStop}%
\bibitem [{\citenamefont {Pakrouski}\ \emph
  {et~al.}(2015{\natexlab{b}})\citenamefont {Pakrouski}, \citenamefont
  {Peterson}, \citenamefont {Jolicoeur}, \citenamefont {Scarola}, \citenamefont
  {Nayak},\ and\ \citenamefont {Troyer}}]{peterson2015}%
  \BibitemOpen
  \bibfield  {author} {\bibinfo {author} {\bibfnamefont {K.}~\bibnamefont
  {Pakrouski}}, \bibinfo {author} {\bibfnamefont {M.~R.}\ \bibnamefont
  {Peterson}}, \bibinfo {author} {\bibfnamefont {T.}~\bibnamefont {Jolicoeur}},
  \bibinfo {author} {\bibfnamefont {V.~W.}\ \bibnamefont {Scarola}}, \bibinfo
  {author} {\bibfnamefont {C.}~\bibnamefont {Nayak}}, \ and\ \bibinfo {author}
  {\bibfnamefont {M.}~\bibnamefont {Troyer}},\ }\href {\doibase
  10.1103/PhysRevX.5.021004} {\bibfield  {journal} {\bibinfo  {journal} {Phys.
  Rev. X}\ }\textbf {\bibinfo {volume} {5}},\ \bibinfo {pages} {021004}
  (\bibinfo {year} {2015}{\natexlab{b}})}\BibitemShut {NoStop}%
\bibitem [{\citenamefont {Estienne}\ \emph {et~al.}(2015)\citenamefont
  {Estienne}, \citenamefont {Regnault},\ and\ \citenamefont
  {Bernevig}}]{estienne2015}%
  \BibitemOpen
  \bibfield  {author} {\bibinfo {author} {\bibfnamefont {B.}~\bibnamefont
  {Estienne}}, \bibinfo {author} {\bibfnamefont {N.}~\bibnamefont {Regnault}},
  \ and\ \bibinfo {author} {\bibfnamefont {B.~A.}\ \bibnamefont {Bernevig}},\
  }\href {\doibase 10.1103/PhysRevLett.114.186801} {\bibfield  {journal}
  {\bibinfo  {journal} {Phys. Rev. Lett.}\ }\textbf {\bibinfo {volume} {114}},\
  \bibinfo {pages} {186801} (\bibinfo {year} {2015})}\BibitemShut {NoStop}%
\bibitem [{fri()}]{friedman1}%
  \BibitemOpen
  \href@noop {} {\bibinfo  {journal} {One exception is the study of
  entanglement entropy of disordered FQH systems by B. A. Friedman, G. C.
  Levine, and D. Luna,
  \href{http://stacks.iop.org/1367-2630/13/i=5/a=055006}{New J. Phys. {\bf 13},
  055006 (2011)}, as well as in the following reference. However, their
  calculations involved a somewhat different partitioning of the system.
  Further, these papers concentrated on extracting the topological term for the
  entanglement entropy, and making contact with experimental results on
  transport, rather than the phase transition, which is the main focus of our
  work}\ }\BibitemShut {NoStop}%
\bibitem [{\citenamefont {Friedman}\ and\ \citenamefont
  {Levine}(2015)}]{friedman2}%
  \BibitemOpen
\bibfield  {journal} {  }\bibfield  {author} {\bibinfo {author} {\bibfnamefont
  {B.~A.}\ \bibnamefont {Friedman}}\ and\ \bibinfo {author} {\bibfnamefont
  {G.~C.}\ \bibnamefont {Levine}},\ }\href {\doibase 10.1142/S0217979215500654}
  {\bibfield  {journal} {\bibinfo  {journal} {Int. J. Mod. Phys B}\ }\textbf
  {\bibinfo {volume} {29}},\ \bibinfo {pages} {1550065} (\bibinfo {year}
  {2015})}\BibitemShut {NoStop}%
\bibitem [{\citenamefont {Yang}\ \emph {et~al.}(2015)\citenamefont {Yang},
  \citenamefont {Chamon}, \citenamefont {Hamma},\ and\ \citenamefont
  {Mucciolo}}]{yangzc}%
  \BibitemOpen
  \bibfield  {author} {\bibinfo {author} {\bibfnamefont {Z.-C.}\ \bibnamefont
  {Yang}}, \bibinfo {author} {\bibfnamefont {C.}~\bibnamefont {Chamon}},
  \bibinfo {author} {\bibfnamefont {A.}~\bibnamefont {Hamma}}, \ and\ \bibinfo
  {author} {\bibfnamefont {E.~R.}\ \bibnamefont {Mucciolo}},\ }\href {\doibase
  10.1103/PhysRevLett.115.267206} {\bibfield  {journal} {\bibinfo  {journal}
  {Phys. Rev. Lett.}\ }\textbf {\bibinfo {volume} {115}},\ \bibinfo {pages}
  {267206} (\bibinfo {year} {2015})}\BibitemShut {NoStop}%
\bibitem [{\citenamefont {Geraedts}\ \emph {et~al.}(2016)\citenamefont
  {Geraedts}, \citenamefont {Nandkishore},\ and\ \citenamefont
  {Regnault}}]{nicolas}%
  \BibitemOpen
  \bibfield  {author} {\bibinfo {author} {\bibfnamefont {S.~D.}\ \bibnamefont
  {Geraedts}}, \bibinfo {author} {\bibfnamefont {R.}~\bibnamefont
  {Nandkishore}}, \ and\ \bibinfo {author} {\bibfnamefont {N.}~\bibnamefont
  {Regnault}},\ }\href {\doibase 10.1103/PhysRevB.93.174202} {\bibfield
  {journal} {\bibinfo  {journal} {Phys. Rev. B}\ }\textbf {\bibinfo {volume}
  {93}},\ \bibinfo {pages} {174202} (\bibinfo {year} {2016})}\BibitemShut
  {NoStop}%
\bibitem [{\citenamefont {Serbyn}\ \emph {et~al.}(2016)\citenamefont {Serbyn},
  \citenamefont {Michailidis}, \citenamefont {Abanin},\ and\ \citenamefont
  {Papi\ifmmode~\acute{c}\else \'{c}\fi{}}}]{powerlawES}%
  \BibitemOpen
  \bibfield  {author} {\bibinfo {author} {\bibfnamefont {M.}~\bibnamefont
  {Serbyn}}, \bibinfo {author} {\bibfnamefont {A.~A.}\ \bibnamefont
  {Michailidis}}, \bibinfo {author} {\bibfnamefont {D.~A.}\ \bibnamefont
  {Abanin}}, \ and\ \bibinfo {author} {\bibfnamefont {Z.}~\bibnamefont
  {Papi\ifmmode~\acute{c}\else \'{c}\fi{}}},\ }\href {\doibase
  10.1103/PhysRevLett.117.160601} {\bibfield  {journal} {\bibinfo  {journal}
  {Phys. Rev. Lett.}\ }\textbf {\bibinfo {volume} {117}},\ \bibinfo {pages}
  {160601} (\bibinfo {year} {2016})}\BibitemShut {NoStop}%
\bibitem [{\citenamefont {Pouranvari}\ \emph {et~al.}(2015)\citenamefont
  {Pouranvari}, \citenamefont {Zhang},\ and\ \citenamefont {Yang}}]{kun}%
  \BibitemOpen
  \bibfield  {author} {\bibinfo {author} {\bibfnamefont {M.}~\bibnamefont
  {Pouranvari}}, \bibinfo {author} {\bibfnamefont {Y.}~\bibnamefont {Zhang}}, \
  and\ \bibinfo {author} {\bibfnamefont {K.}~\bibnamefont {Yang}},\ }\href
  {\doibase 10.1155/2015/397630} {\bibfield  {journal} {\bibinfo  {journal}
  {Adv. Condens. Matter Phys.}\ }\textbf {\bibinfo {volume} {2015}},\ \bibinfo
  {pages} {397630} (\bibinfo {year} {2015})}\BibitemShut {NoStop}%
\bibitem [{\citenamefont {Chandran}\ \emph {et~al.}()\citenamefont {Chandran},
  \citenamefont {Laumann},\ and\ \citenamefont {Oganesyan}}]{Chandran}%
  \BibitemOpen
  \bibfield  {author} {\bibinfo {author} {\bibfnamefont {A.}~\bibnamefont
  {Chandran}}, \bibinfo {author} {\bibfnamefont {C.~R.}\ \bibnamefont
  {Laumann}}, \ and\ \bibinfo {author} {\bibfnamefont {V.}~\bibnamefont
  {Oganesyan}},\ }\href@noop {} {\ }\Eprint {http://arxiv.org/abs/1509.04285}
  {arXiv:1509.04285} \BibitemShut {NoStop}%
\bibitem [{new()}]{newref}%
  \BibitemOpen
  \href@noop {} {\bibinfo  {journal} {FSS requires that the system size $L$ and
  the correlation length $\xi$ be both much larger than the microscopic length
  ($\ell_B$, the magnetic length). However, it allows any value of the ratio
  $L/ \xi$, and is therefore less stringent in its requirements than, {\it
  e.g.,} asymptotic formulas for entanglement entropy, which require $L \gg
  \xi$}\ }\BibitemShut {NoStop}%
\bibitem [{\citenamefont {Chayes}\ \emph {et~al.}(1986)\citenamefont {Chayes},
  \citenamefont {Chayes}, \citenamefont {Fisher},\ and\ \citenamefont
  {Spencer}}]{Chayes86}%
  \BibitemOpen
\bibfield  {journal} {  }\bibfield  {author} {\bibinfo {author} {\bibfnamefont
  {J.~T.}\ \bibnamefont {Chayes}}, \bibinfo {author} {\bibfnamefont
  {L.}~\bibnamefont {Chayes}}, \bibinfo {author} {\bibfnamefont {D.~S.}\
  \bibnamefont {Fisher}}, \ and\ \bibinfo {author} {\bibfnamefont
  {T.}~\bibnamefont {Spencer}},\ }\href {\doibase 10.1103/PhysRevLett.57.2999}
  {\bibfield  {journal} {\bibinfo  {journal} {Phys. Rev. Lett.}\ }\textbf
  {\bibinfo {volume} {57}},\ \bibinfo {pages} {2999} (\bibinfo {year}
  {1986})}\BibitemShut {NoStop}%
\bibitem [{\citenamefont {Harris}(1974)}]{harris}%
  \BibitemOpen
  \bibfield  {author} {\bibinfo {author} {\bibfnamefont {A.~B.}\ \bibnamefont
  {Harris}},\ }\href {http://stacks.iop.org/0022-3719/7/i=17/a=018} {\bibfield
  {journal} {\bibinfo  {journal} {J. Phys. C}\ }\textbf {\bibinfo {volume}
  {7}},\ \bibinfo {pages} {3082} (\bibinfo {year} {1974})}\BibitemShut
  {NoStop}%
\bibitem [{\citenamefont {Mott}(1976)}]{Mott}%
  \BibitemOpen
  \bibfield  {author} {\bibinfo {author} {\bibfnamefont {N.~F.}\ \bibnamefont
  {Mott}},\ }\href@noop {} {\bibfield  {journal} {\bibinfo  {journal} {Commun.
  Phys.}\ }\textbf {\bibinfo {volume} {1}},\ \bibinfo {pages} {203} (\bibinfo
  {year} {1976})}\BibitemShut {NoStop}%
\bibitem [{\citenamefont {Kj\"all}\ \emph {et~al.}(2014)\citenamefont
  {Kj\"all}, \citenamefont {Bardarson},\ and\ \citenamefont
  {Pollmann}}]{pollman}%
  \BibitemOpen
  \bibfield  {author} {\bibinfo {author} {\bibfnamefont {J.~A.}\ \bibnamefont
  {Kj\"all}}, \bibinfo {author} {\bibfnamefont {J.~H.}\ \bibnamefont
  {Bardarson}}, \ and\ \bibinfo {author} {\bibfnamefont {F.}~\bibnamefont
  {Pollmann}},\ }\href {\doibase 10.1103/PhysRevLett.113.107204} {\bibfield
  {journal} {\bibinfo  {journal} {Phys. Rev. Lett.}\ }\textbf {\bibinfo
  {volume} {113}},\ \bibinfo {pages} {107204} (\bibinfo {year}
  {2014})}\BibitemShut {NoStop}%
\bibitem [{\citenamefont {Luitz}\ \emph {et~al.}(2015)\citenamefont {Luitz},
  \citenamefont {Laflorencie},\ and\ \citenamefont {Alet}}]{alet}%
  \BibitemOpen
  \bibfield  {author} {\bibinfo {author} {\bibfnamefont {D.~J.}\ \bibnamefont
  {Luitz}}, \bibinfo {author} {\bibfnamefont {N.}~\bibnamefont {Laflorencie}},
  \ and\ \bibinfo {author} {\bibfnamefont {F.}~\bibnamefont {Alet}},\ }\href
  {\doibase 10.1103/PhysRevB.91.081103} {\bibfield  {journal} {\bibinfo
  {journal} {Phys. Rev. B}\ }\textbf {\bibinfo {volume} {91}},\ \bibinfo
  {pages} {081103(R)} (\bibinfo {year} {2015})}\BibitemShut {NoStop}%
\bibitem [{\citenamefont {Haldane}(1983)}]{haldaneVm}%
  \BibitemOpen
  \bibfield  {author} {\bibinfo {author} {\bibfnamefont {F.~D.~M.}\
  \bibnamefont {Haldane}},\ }\href {\doibase 10.1103/PhysRevLett.51.605}
  {\bibfield  {journal} {\bibinfo  {journal} {Phys. Rev. Lett.}\ }\textbf
  {\bibinfo {volume} {51}},\ \bibinfo {pages} {605} (\bibinfo {year}
  {1983})}\BibitemShut {NoStop}%
\bibitem [{V1_()}]{V1_note}%
  \BibitemOpen
  \href@noop {} {\bibinfo  {journal} {We normalize the pseudopotential such
  that the energy scale of the two-particle problem is unity}\ }\BibitemShut
  {NoStop}%
\bibitem [{\citenamefont {Laughlin}(1983)}]{laughlin}%
  \BibitemOpen
\bibfield  {journal} {  }\bibfield  {author} {\bibinfo {author} {\bibfnamefont
  {R.~B.}\ \bibnamefont {Laughlin}},\ }\href {\doibase
  10.1103/PhysRevLett.50.1395} {\bibfield  {journal} {\bibinfo  {journal}
  {Phys. Rev. Lett.}\ }\textbf {\bibinfo {volume} {50}},\ \bibinfo {pages}
  {1395} (\bibinfo {year} {1983})}\BibitemShut {NoStop}%
\bibitem [{sys()}]{systemsize}%
  \BibitemOpen
  \href@noop {} {\bibinfo  {journal} {In this work, the largest system size
  that we can reach by exact diagonalization is $N=10$ electrons, whose Hilbert
  space dimension is $30045015$. This system size is smaller than the limit in
  clean systems due to the breaking of translation invariance by disorder.
  Considering that we often need at least hundreds of samples to do the
  ensemble average, we focus on $N\leq 9$ electrons (with Hilbert space
  dimension up to $4686825$) in most of our calculations to finish the
  computation in a reasonable time}\ }\BibitemShut {NoStop}%
\bibitem [{\citenamefont {Wen}\ and\ \citenamefont {Niu}(1990)}]{wen}%
  \BibitemOpen
\bibfield  {journal} {  }\bibfield  {author} {\bibinfo {author} {\bibfnamefont
  {X.~G.}\ \bibnamefont {Wen}}\ and\ \bibinfo {author} {\bibfnamefont
  {Q.}~\bibnamefont {Niu}},\ }\href {\doibase 10.1103/PhysRevB.41.9377}
  {\bibfield  {journal} {\bibinfo  {journal} {Phys. Rev. B}\ }\textbf {\bibinfo
  {volume} {41}},\ \bibinfo {pages} {9377} (\bibinfo {year}
  {1990})}\BibitemShut {NoStop}%
\bibitem [{sup()}]{supple}%
  \BibitemOpen
  \href@noop {} {\bibinfo  {journal} {See the Supplemental Material for more
  numerical data of the ground-state entanglement entropy and entanglement
  spectrum}\ }\BibitemShut {NoStop}%
\bibitem [{dsd()}]{dsdw_note}%
  \BibitemOpen
\bibfield  {journal} {  }\href@noop {} {\bibinfo  {journal} {The disorder
  configuration is fixed when we compute $S(\overline{\rho})|_{W+\Delta W}$ and
  $S(\overline{\rho})|_W$. Only the magnitude of $W$ is changed by a small
  percentage. We use $\Delta W=0.001W$, though $\Delta W=0.01W$ gives almost
  the same results}\ }\BibitemShut {NoStop}%
\bibitem [{\citenamefont {Dong}\ \emph {et~al.}(2008)\citenamefont {Dong},
  \citenamefont {Fradkin}, \citenamefont {Leigh},\ and\ \citenamefont
  {Nowling}}]{dong2008}%
  \BibitemOpen
\bibfield  {journal} {  }\bibfield  {author} {\bibinfo {author} {\bibfnamefont
  {S.}~\bibnamefont {Dong}}, \bibinfo {author} {\bibfnamefont {E.}~\bibnamefont
  {Fradkin}}, \bibinfo {author} {\bibfnamefont {R.~G.}\ \bibnamefont {Leigh}},
  \ and\ \bibinfo {author} {\bibfnamefont {S.}~\bibnamefont {Nowling}},\ }\href
  {http://stacks.iop.org/1126-6708/2008/i=05/a=016} {\bibfield  {journal}
  {\bibinfo  {journal} {J. High Energy Phys.}\ }\textbf {\bibinfo {volume}
  {2008}},\ \bibinfo {pages} {016} (\bibinfo {year} {2008})}\BibitemShut
  {NoStop}%
\bibitem [{\citenamefont {Zhang}\ \emph {et~al.}(2012)\citenamefont {Zhang},
  \citenamefont {Grover}, \citenamefont {Turner}, \citenamefont {Oshikawa},\
  and\ \citenamefont {Vishwanath}}]{zhangyi}%
  \BibitemOpen
  \bibfield  {author} {\bibinfo {author} {\bibfnamefont {Y.}~\bibnamefont
  {Zhang}}, \bibinfo {author} {\bibfnamefont {T.}~\bibnamefont {Grover}},
  \bibinfo {author} {\bibfnamefont {A.}~\bibnamefont {Turner}}, \bibinfo
  {author} {\bibfnamefont {M.}~\bibnamefont {Oshikawa}}, \ and\ \bibinfo
  {author} {\bibfnamefont {A.}~\bibnamefont {Vishwanath}},\ }\href {\doibase
  10.1103/PhysRevB.85.235151} {\bibfield  {journal} {\bibinfo  {journal} {Phys.
  Rev. B}\ }\textbf {\bibinfo {volume} {85}},\ \bibinfo {pages} {235151}
  (\bibinfo {year} {2012})}\BibitemShut {NoStop}%
\bibitem [{\citenamefont {Zhu}\ \emph {et~al.}(2013)\citenamefont {Zhu},
  \citenamefont {Sheng},\ and\ \citenamefont {Haldane}}]{wei1}%
  \BibitemOpen
  \bibfield  {author} {\bibinfo {author} {\bibfnamefont {W.}~\bibnamefont
  {Zhu}}, \bibinfo {author} {\bibfnamefont {D.~N.}\ \bibnamefont {Sheng}}, \
  and\ \bibinfo {author} {\bibfnamefont {F.~D.~M.}\ \bibnamefont {Haldane}},\
  }\href {\doibase 10.1103/PhysRevB.88.035122} {\bibfield  {journal} {\bibinfo
  {journal} {Phys. Rev. B}\ }\textbf {\bibinfo {volume} {88}},\ \bibinfo
  {pages} {035122} (\bibinfo {year} {2013})}\BibitemShut {NoStop}%
\bibitem [{\citenamefont {Zhu}\ \emph {et~al.}(2014)\citenamefont {Zhu},
  \citenamefont {Gong}, \citenamefont {Haldane},\ and\ \citenamefont
  {Sheng}}]{wei2}%
  \BibitemOpen
  \bibfield  {author} {\bibinfo {author} {\bibfnamefont {W.}~\bibnamefont
  {Zhu}}, \bibinfo {author} {\bibfnamefont {S.~S.}\ \bibnamefont {Gong}},
  \bibinfo {author} {\bibfnamefont {F.~D.~M.}\ \bibnamefont {Haldane}}, \ and\
  \bibinfo {author} {\bibfnamefont {D.~N.}\ \bibnamefont {Sheng}},\ }\href
  {\doibase 10.1103/PhysRevLett.112.096803} {\bibfield  {journal} {\bibinfo
  {journal} {Phys. Rev. Lett.}\ }\textbf {\bibinfo {volume} {112}},\ \bibinfo
  {pages} {096803} (\bibinfo {year} {2014})}\BibitemShut {NoStop}%
\bibitem [{cle()}]{cleanmes}%
  \BibitemOpen
  \href@noop {} {\bibinfo  {journal} {In clean samples ($W=0$) with translation
  invariance, there are three minimally entangled states with different
  momenta, corresponding to the three types of quasiparticles of the Laughlin
  state}\ }\BibitemShut {NoStop}%
\bibitem [{\citenamefont {Liu}\ \emph {et~al.}(2012{\natexlab{b}})\citenamefont
  {Liu}, \citenamefont {Bergholtz}, \citenamefont {Fan},\ and\ \citenamefont
  {L\"auchli}}]{zliu}%
  \BibitemOpen
\bibfield  {journal} {  }\bibfield  {author} {\bibinfo {author} {\bibfnamefont
  {Z.}~\bibnamefont {Liu}}, \bibinfo {author} {\bibfnamefont {E.~J.}\
  \bibnamefont {Bergholtz}}, \bibinfo {author} {\bibfnamefont {H.}~\bibnamefont
  {Fan}}, \ and\ \bibinfo {author} {\bibfnamefont {A.~M.}\ \bibnamefont
  {L\"auchli}},\ }\href {\doibase 10.1103/PhysRevB.85.045119} {\bibfield
  {journal} {\bibinfo  {journal} {Phys. Rev. B}\ }\textbf {\bibinfo {volume}
  {85}},\ \bibinfo {pages} {045119} (\bibinfo {year}
  {2012}{\natexlab{b}})}\BibitemShut {NoStop}%
\bibitem [{\citenamefont {Prodan}\ \emph {et~al.}(2010)\citenamefont {Prodan},
  \citenamefont {Hughes},\ and\ \citenamefont {Bernevig}}]{Prodan}%
  \BibitemOpen
  \bibfield  {author} {\bibinfo {author} {\bibfnamefont {E.}~\bibnamefont
  {Prodan}}, \bibinfo {author} {\bibfnamefont {T.~L.}\ \bibnamefont {Hughes}},
  \ and\ \bibinfo {author} {\bibfnamefont {B.~A.}\ \bibnamefont {Bernevig}},\
  }\href {\doibase 10.1103/PhysRevLett.105.115501} {\bibfield  {journal}
  {\bibinfo  {journal} {Phys. Rev. Lett.}\ }\textbf {\bibinfo {volume} {105}},\
  \bibinfo {pages} {115501} (\bibinfo {year} {2010})}\BibitemShut {NoStop}%
\bibitem [{win()}]{window_note}%
  \BibitemOpen
  \href@noop {} {\bibinfo  {journal} {The division of windows depends on the
  best matching between $\overline{P}(s)$ in each window and the candidate
  level statistics (GUE, semi-Poisson, or Poisson). We choose
  $\xi\in(0,10],(10,20]$ and $(20,40]$ at $W=0.4,0.6,1,10$, and
  $\xi\in(0,15],(15,20]$ and $(20,40]$ at $W=100,\infty$}\ }\BibitemShut
  {NoStop}%
\bibitem [{n<4()}]{n<40}%
  \BibitemOpen
\bibfield  {journal} {  }\href@noop {} {\bibinfo  {journal} {We compare the
  number of reliable ES levels below $\xi=40$, i.e.,
  $n_{\leq40}=\int_{0}^{40}\langle\overline{D}(\xi)\rangle d\xi$, with the
  total number of ES levels $n_{\rm{tot}}=\tbinom{\lceil
  N_\phi/2\rfloor}{\lceil N/2\rfloor}$. For $N=9$ electrons, we have
  $n_{\rm{tot}}=715$ and $n_{\leq40}=715,715,715,683,468$ and $126$ at
  $W=0.4,0.6,1,10,100$ and $\infty$, respectively}\ }\BibitemShut {NoStop}%
\bibitem [{\citenamefont {Liu}\ and\ \citenamefont {Bhatt}()}]{zhao-bhatt}%
  \BibitemOpen
\bibfield  {journal} {  }\bibfield  {author} {\bibinfo {author} {\bibfnamefont
  {Z.}~\bibnamefont {Liu}}\ and\ \bibinfo {author} {\bibfnamefont {R.~N.}\
  \bibnamefont {Bhatt}},\ }\href@noop {} {\bibinfo  {journal} {to be
  published}\ }\BibitemShut {NoStop}%
\bibitem [{\citenamefont {Shahar}\ \emph {et~al.}(1995)\citenamefont {Shahar},
  \citenamefont {Tsui}, \citenamefont {Shayegan}, \citenamefont {Bhatt},\ and\
  \citenamefont {Cunningham}}]{shahar}%
  \BibitemOpen
\bibfield  {journal} {  }\bibfield  {author} {\bibinfo {author} {\bibfnamefont
  {D.}~\bibnamefont {Shahar}}, \bibinfo {author} {\bibfnamefont {D.~C.}\
  \bibnamefont {Tsui}}, \bibinfo {author} {\bibfnamefont {M.}~\bibnamefont
  {Shayegan}}, \bibinfo {author} {\bibfnamefont {R.~N.}\ \bibnamefont {Bhatt}},
  \ and\ \bibinfo {author} {\bibfnamefont {J.~E.}\ \bibnamefont {Cunningham}},\
  }\href {\doibase 10.1103/PhysRevLett.74.4511} {\bibfield  {journal} {\bibinfo
   {journal} {Phys. Rev. Lett.}\ }\textbf {\bibinfo {volume} {74}},\ \bibinfo
  {pages} {4511} (\bibinfo {year} {1995})}\BibitemShut {NoStop}%
\bibitem [{\citenamefont {Lee}\ \emph {et~al.}(1992)\citenamefont {Lee},
  \citenamefont {Kivelson},\ and\ \citenamefont {Zhang}}]{kivelson}%
  \BibitemOpen
  \bibfield  {author} {\bibinfo {author} {\bibfnamefont {D.-H.}\ \bibnamefont
  {Lee}}, \bibinfo {author} {\bibfnamefont {S.}~\bibnamefont {Kivelson}}, \
  and\ \bibinfo {author} {\bibfnamefont {S.-C.}\ \bibnamefont {Zhang}},\ }\href
  {\doibase 10.1103/PhysRevLett.68.2386} {\bibfield  {journal} {\bibinfo
  {journal} {Phys. Rev. Lett.}\ }\textbf {\bibinfo {volume} {68}},\ \bibinfo
  {pages} {2386} (\bibinfo {year} {1992})}\BibitemShut {NoStop}%
\bibitem [{\citenamefont {Huo}\ \emph {et~al.}(1993)\citenamefont {Huo},
  \citenamefont {Hetzel},\ and\ \citenamefont {Bhatt}}]{huohetzelbhatt}%
  \BibitemOpen
  \bibfield  {author} {\bibinfo {author} {\bibfnamefont {Y.}~\bibnamefont
  {Huo}}, \bibinfo {author} {\bibfnamefont {R.~E.}\ \bibnamefont {Hetzel}}, \
  and\ \bibinfo {author} {\bibfnamefont {R.~N.}\ \bibnamefont {Bhatt}},\ }\href
  {\doibase 10.1103/PhysRevLett.70.481} {\bibfield  {journal} {\bibinfo
  {journal} {Phys. Rev. Lett.}\ }\textbf {\bibinfo {volume} {70}},\ \bibinfo
  {pages} {481} (\bibinfo {year} {1993})}\BibitemShut {NoStop}%
\end{thebibliography}%

%\clearpage
\section*{Supplemental Material}

In this Supplemental Material, we provide more numerical data for the ground-state entanglement entropy and entanglement spectrum.

\subsection*{Ground-state entanglement entropy}

\begin{figure*}
\centerline{\includegraphics[width=\linewidth]{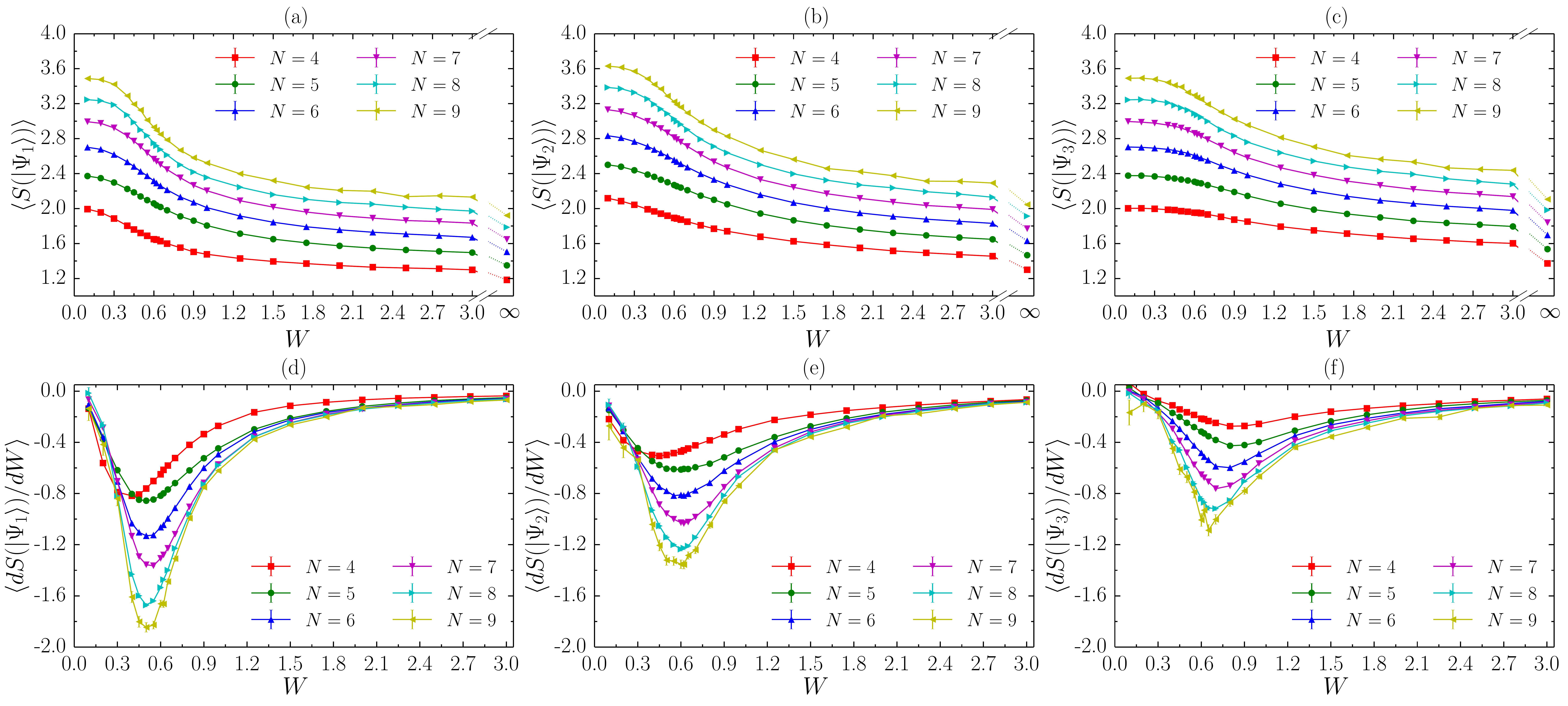}}
\caption{$\langle S(|\Psi_i\rangle)\rangle$ and $\langle dS(|\Psi_i\rangle)/dW\rangle$ for (a,d) $|\Psi_1\rangle$, (b,e) $|\Psi_2\rangle$ and (c,f) $|\Psi_3\rangle$, where $|\Psi_1\rangle$, $|\Psi_2\rangle$ and $|\Psi_3\rangle$ are the three states with ascending energies in the ground-state manifold. Here we averaged $20000$ samples for $N=4-7$, $5000$ samples for $N=8$, and $800$ samples for $N=9$ electrons. The data at $W=\infty$, i.e., the noninteracting limit are also given.}
\label{Spsi}
\end{figure*}

\begin{figure}
\centerline{\includegraphics[width=\linewidth]{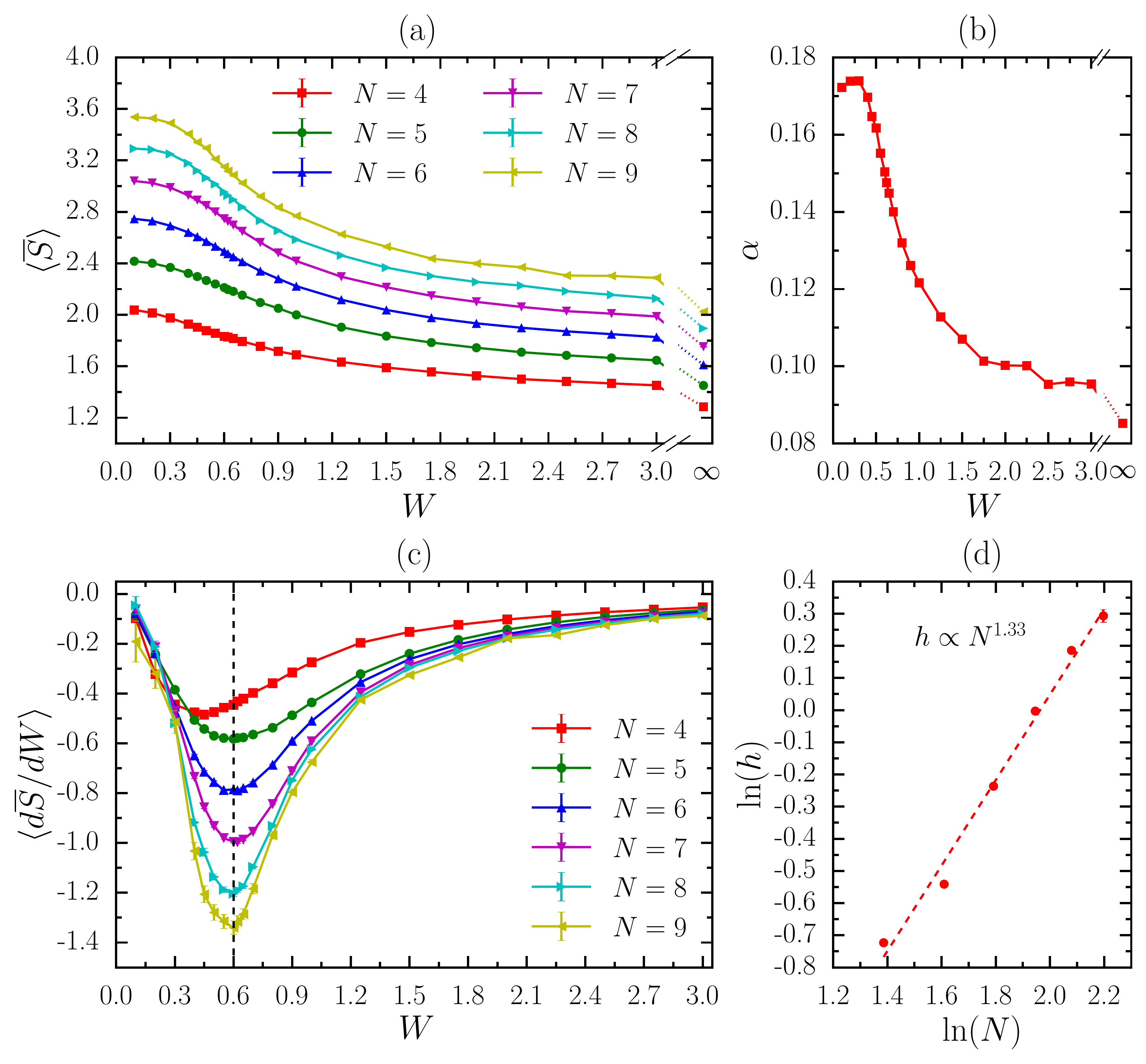}}
\caption{We measure the ground-state entanglement by $\overline{S}=\frac{1}{3}\sum_{i=1}^3 S(|\Psi_i\rangle)$. (a) $\langle\overline{S}\rangle$, (b) the entanglement density $\alpha$, and (c) $\langle d\overline{S}/dW\rangle$ versus the disorder strength $W$. (d) The depth $h$ of $\langle d\overline{S}/dW\rangle$ versus the number of electrons $N$ on a double logarithmic plot. The linear fit (dashed line) shows $h\propto N^{1.33}$. Here we averaged $20000$ samples for $N=4-7$, $5000$ samples for $N=8$, and $800$ samples for $N=9$ electrons. The data at $W=\infty$, i.e., the noninteracting limit are also given in (a) and (b).}
\label{Sbar}
\end{figure}

In the main text, we have discussed the ground-state entanglement entropy $S(\overline{\rho})$ obtained by averaging the density matrices of the three ground states, i.e., $\overline{\rho}=\frac{1}{3}\sum_{i=1}^3|\Psi_i\rangle\langle\Psi_i|$. Now we compute the corresponding result $S(|\Psi_i\rangle)$ and its derivative $dS(|\Psi_i\rangle)/dW$ of the three individual states.
The sample-averaged results are shown in Fig.~\ref{Spsi}. The data of three individual states have some differences, but are qualitatively the same: for all of them, the entanglement decreases with $W$, and the derivative with respect to $W$ has a single minimum that becomes deeper for larger system sizes. For the finite systems that we have studied, the location of the minimum does depend somewhat on the individual states, but the value does not deviate much from $W=0.6$. To incorporate the effects of all of the three states, we compute the mean $\overline{S}=\frac{1}{3}\sum_{i=1}^3 S(|\Psi_i\rangle)$. This is an alternative averaging method to the one ($\overline{\rho}=\frac{1}{3}\sum_{i=1}^3|\Psi_i\rangle\langle\Psi_i|$) that we use in the main text. The sample-averaged results are shown in Fig.~\ref{Sbar}. The minimum of $\langle d\overline{S}/dW\rangle$ is located at $W\approx0.6$ for $N=5-9$ electrons [Fig.~\ref{Sbar}(c)], and its depth diverges as $h\propto N^{1.33}$ with the system size [Fig.~\ref{Sbar}(d)]. The scaling $d\overline{S}/dW\propto N^{\frac{1}{2}+\frac{1}{2\nu}}f'[N^{\frac{1}{2\nu}}(W-W_c)]$ suggests $\nu\approx 0.6$. $\langle\overline{S}\rangle$ agrees with an area law at all $W$'s, and the entanglement density starts to drop at $W\approx0.4$ [Fig.~\ref{Sbar}(b)]. All of these results are very similar to those shown in Figs.~1 and 2 in the main text. This means both averaging methods, i.e., $\overline{\rho}=\frac{1}{3}\sum_{i=1}^3|\Psi_i\rangle\langle\Psi_i|$ and $\overline{S}=\frac{1}{3}\sum_{i=1}^3 S(|\Psi_i\rangle)$, can identify the ground-state phase transitions and give the same critical $W$. However, we observe larger finite-size effects of $h$ and error bars in $\langle d\overline{S}/dW\rangle$ (especially at small $W$).

\subsection*{Ground-state entanglement spectrum (ES)}

\begin{figure}
\centerline{\includegraphics[width=\linewidth]{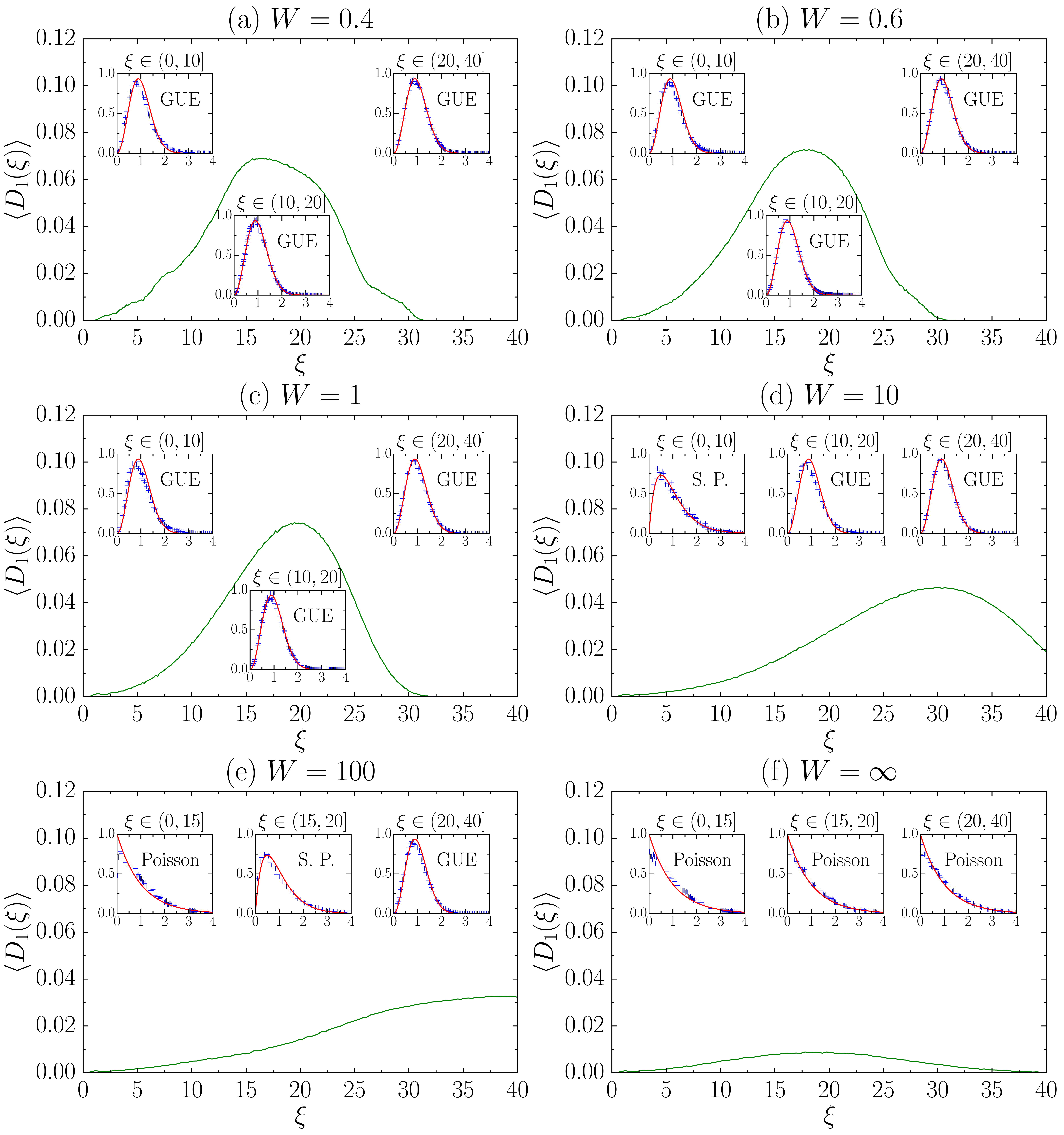}}
\caption{The sample-averaged DOS $\langle D_1(\xi)\rangle$ and the level-spacing distribution $P_1(s)$ of the ground-state ES below $\xi=40$ for $|\Psi_1\rangle$ of $N=9$ electrons at (a) $W=0.4$, (b) $W=0.6$, (c) $W=1$, (d) $W=10$, (e) $W=100$ and (f) $W=\infty$. At each $W$, we choose three windows to compute $P_1(s)$, plotted versus $s$ in the insets. The blue crosses correspond to numerical data, while the red lines give the theoretical prediction for the Gaussian unitary ensemble (GUE), semi-Poisson (S.~P.) and the Poisson distribution, for which $P(s)=\frac{32}{\pi^2}s^2 e^{-\frac{4}{\pi}s^2}$, $P(s)=4se^{-2s}$ and $P(s)=e^{-s}$, respectively. Data from $800$ realizations of disorder.
}
\label{oespsi1}
\end{figure}

In the main text, we consider the density of states (DOS) $\overline{D}(\xi)$ and level statistics $\overline{P}(s)$ of the ES averaged over three ground states. We find that the results of each individual state are almost the same as those obtained by averaging over three ground states, which justifies the procedures of doing an average. Here, we demonstrate the results [$D_1(\xi)$ and $P_1(s)$] of $|\Psi_1\rangle$ for completeness (Fig.~\ref{oespsi1}). The results for $|\Psi_2\rangle$ and $|\Psi_3\rangle$ are almost the same as $|\Psi_1\rangle$, thus we do not show them here.

\begin{figure}
\centerline{\includegraphics[width=\linewidth]{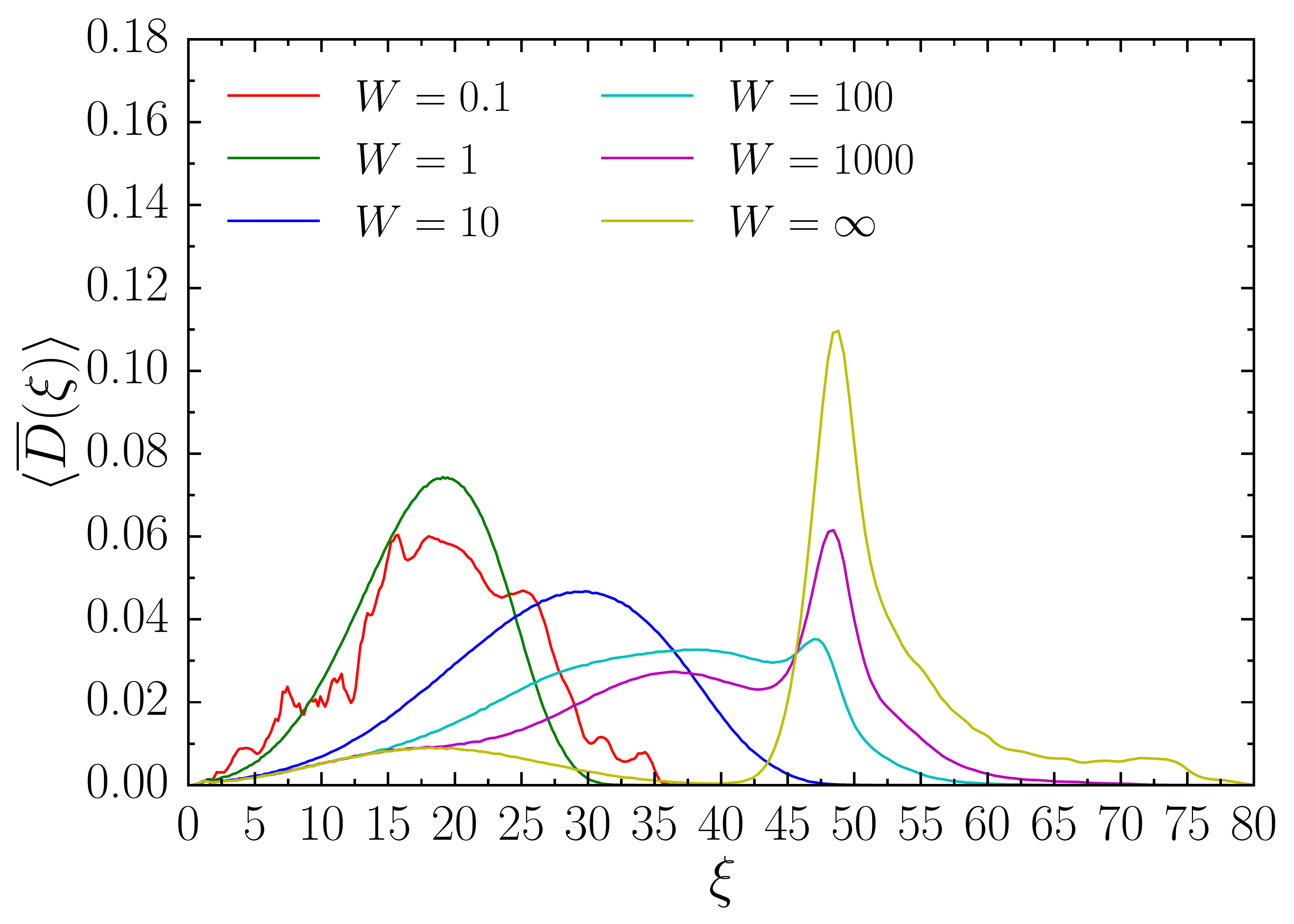}}
\caption{The sample-averaged DOS $\langle\overline{D}(\xi)\rangle$ of the ground-state ES of $N=9$ electrons at $W=0.1,1,10,100,1000$ and $\infty$. $\langle\overline{D}(\xi)\rangle$ is averaged over the three ground states using $800$ samples.}
\label{oesdos}
\end{figure}

We should also consider the problem of numerical noise in the ES obtained by singular value decomposition of the many-body eigenstates. The machine precision for double precision variables is $2^{-53}$. This implies that those singular values $\sqrt{\xi}$ below $2^{-53}$ have the danger to be ruined by the numerical noise, which corresponds to $\xi=-\ln2^{-53\times 2}\approx 73.5$ in the ES. Considering that the entries of the many-body eigenstates are complex numbers (two double precision variables) in our systems and the many-body eigenstates themselves also contain numerical error from Lanczos iterations, the numerical noise in the ES may appear at lower $\xi$. In order to detect the critical $\xi$ at which the machine precision problem starts to dominate, we check the DOS $\overline{D}(\xi)$ of the ES at different disorder strengths. We expect that the ES levels generated by numerical noise always assemble around the same energy. This will correspond to a peak in the DOS that does not move with the change of disorder strength. In Fig.~\ref{oesdos}, we indeed observe such a situation deeply in the localized phase. There is always a peak around $\xi\approx50$ that does not move for $W=100,1000$ and $\infty$, meaning that the machine precision problem has occured at these disorder strengths. Therefore, we only focus on those ES levels with $\xi\leq40$ for safety. 

\end{document}